\documentclass[final,times,twocolumn,5p]{elsarticle}
\usepackage{amssymb}
\usepackage[utf8]{inputenc}
\usepackage[T1]{fontenc}
\usepackage{graphicx}
\usepackage{url}
\usepackage{tabularx}
\usepackage[caption=false]{subfig}
\usepackage[shadow,backgroundcolor=orange!40]{todonotes}

\biboptions{square}
\journal{Future Generation Computer Systems}

\begin{document}
\begin{frontmatter}


\title{Elastic Business Process Management:\\State of the Art and Open Challenges for BPM in the Cloud}

\author[tuv]{Stefan Schulte}
\author[jmu]{Christian Janiesch}
\author[unsw]{Srikumar Venugopal}
\author[nicta,unsw]{Ingo Weber}
\author[tuv]{Philipp Hoenisch}

\address[tuv]{Vienna University of Technology, Vienna, Austria, \{s.schulte $|$ p.hoenisch\}@infosys.tuwien.ac.at}
\address[jmu]{Julius Maximilian University of W\"{u}rzburg, W\"{u}rzburg, Germany, christian.janiesch@uni-wuerzburg.de}
\address[unsw]{The University of New South Wales, Sydney, Australia, srikumarv@cse.unsw.edu.au}
\address[nicta]{NICTA \& The University of New South Wales, Sydney, Australia, ingo.weber@nicta.com.au}
\begin{abstract}

With the advent of cloud computing, organizations are nowadays able to react rapidly to changing demands for computational resources. Not only individual applications can be hosted on virtual cloud infrastructures, but also complete business processes. This allows the realization of so-called elastic processes, i.e., processes which are carried out using elastic cloud resources. Despite the manifold benefits of elastic processes, there is still a lack of solutions supporting them.

In this paper, we identify the state of the art of elastic Business Process Management with a focus on infrastructural challenges. We conceptualize an architecture for an elastic Business Process Management System and discuss existing work on scheduling, resource allocation, monitoring, decentralized coordination, and state management for elastic processes. Furthermore, we present two representative elastic Business Process Management Systems which are intended to counter these challenges. Based on our findings, we identify open issues and outline possible research directions for the realization of elastic processes and elastic Business Process Management.
\end{abstract}

\begin{keyword}
Elastic BPM \sep Elastic Processes \sep Cloud Computing \sep Business Process Management 


\vspace{1cm}
\noindent
\fbox{\parbox{2\linewidth}{
NOTICE: this is the author’s version of a work that was accepted for publication in Future Generation Computer Systems. Changes resulting from the publishing process, such as peer review, editing, corrections, structural formatting, and other quality control mechanisms may not be reflected in this document. Changes may have been made to this work since it was submitted for publication. A definitive version was subsequently published in Future Generation Computer Systems as:\\ \textbf{S. Schulte, C. Janiesch, S. Venugopal, I. Weber, and P. Hoenisch (2015). Elastic Business Process Management: State of the Art and Open Challenges for BPM in the Cloud. Future Generation Computer Systems, Volume NN, Number N, NN-NN., \url{http://dx.doi.org/10.1016/j.future.2014.09.005}}.
}}
\end{keyword}

\end{frontmatter}

\section{Introduction}
\label{sec:intro}
Business Process Management (BPM) enables flexible and individualistic composition and execution of services as opposed to hard-coded workflows in off-the-shelf software. Scalability of computational resources for these processes is an important challenge to realize in the next generation of inter-organizational processes \cite{breu13}. In a traditional architecture, resources for human and service tasks within these processes had to be scheduled in advance and were sized so that they were able to allow for processing of tasks even at peak times. However, in non peak-times, resources would not be utilized, leading to \emph{overprovisioning}. On the other hand, if resources were provided at a level that can only cover part of the processes' resource demands during peak times, processes can either not be carried out (\emph{underprovisioning}) or provide low Quality of Service (QoS) \cite{Armbrust10}. 

Cloud computing has changed how computing and storage services are provisioned. Through elastic, virtualized infrastructures, it is possible to lease and release the needed resources in an \emph{on-demand, utility-like} fashion, with billing according to use (\emph{metered service} or \emph{pay-per-use}), and to scale the computing infrastructure up and down rapidly (\emph{rapid elasticity}) \cite{Armbrust10}. However, we found that as of now, BPM cloud offerings solely focus on providing on-demand platforms and software over the internet in addition to selling traditional software licenses for server installations. Currently, there is very little information on resource elasticity of these offerings or even BPM software to run your own cloud-enabled BPM system.

Scientific workflows (SWFs) \cite{bark08} and corporate data transformation processes, e.g. in master data management or logistics, among others can benefit from such systems. Both have heterogeneous yet mostly heavy computational and storage requirements. This is due to the fact that several steps of calculations and data transformations have to be made in a certain order. Simply providing a Web interface for process design and enactment does not suffice in these cases. It is important to also have access to elastic cloud resources for process enactment \cite{voec11}.

We believe this lack of elasticity support is due to the fact that for BPM, recent advances in cloud-based application provisioning cannot be transferred unmodified but have to be adapted to ensure consistent behavior when executing processes. For example, business processes carry their own state and may execute over a timeframe that can be longer than the life of a Virtual Machine (VM) provisioned in a given scenario. Hence, certain adjustments have to be made when executing processes. In addition, BPM knowledge, available through monitoring services, can be used to improve the resource prediction of current state of the art Infrastructure as a Service (IaaS) auto-scalers. This enables efficient leasing and releasing of computational resources to support execution of process tasks within time and cost constraints.

In this paper, we outline recent advances in BPM and cloud computing that have brought these two worlds closer together. We subsume them under the term \textit{elastic BPM} -- correspondingly, the entities we are discussing are \emph{elastic processes} \cite{dustdar11}. Our focus is on the part of process enactment, as opposed to other phases of the BPM lifecycle such as process design or process evaluation. There are at least two layers to consider when talking about elastic BPM: a business layer, covering all aspects which cannot be implemented to run automatically through a Business Process Management System (BPMS), and an infrastructure layer, covering all aspects of BPM execution. In this paper, we focus primarily on the infrastructure challenges posed and opportunities given by cloud computing and BPM.

We conceive different infrastructure challenges as separate threads or portions of a high level elastic BPM architecture which have interdependencies and intertwine. Following the common states of process enactment, initially there is the challenge of process scheduling: the ordering and execution of processes and their tasks on available resources. Alongside any execution, there is the monitoring of audit data, which presently involves input from and output to third party data streams in real-time. The automatic processing of individual tasks consumes computational resources. The cost-efficient provisioning and deprovisioning of virtualized infrastructure has to go hand in hand with process enactment and the process instances' states. Finally, on a lower level including storage and state management of the applications serving the process, consistency issues have to be dealt with. These topics form the basis of our argument for a coalescence of BPM and cloud computing to allow both worlds to benefit from each other.

The paper provides first hand evidence of the state of the art of performing BPM on cloud infrastructures. In the course of designing and developing our research prototypes for elastic BPM, presented in a later section, we have encountered several obstacles for BPM in the cloud. We have compiled these into five distinct challenges which need to be countered. Concepts and prototypes for some aspects can be found in our research implementations while some other aspects still require further research. For the latter, we have fleshed out concrete research directions for the BPM and cloud communities to investigate, so as to enable truly elastic BPM.

This paper is structured as follows. First, we briefly present background work on BPM and cloud computing (Section~\ref{sec:background}). Then we introduce a BPMS for elastic processes (Section~\ref{sec:architecture}) and an overview of the current state of the art regarding the identified infrastructural challenges (Section~\ref{sec:challenges}). We underpin our main section on elastic BPM by describing two prototypical research prototypes which have been used to evaluate the benefits of joint BPM and cloud computing realizations (Section~\ref{sec:realisations}). Finally we summarize the findings and systematize future research directions in Sections~\ref{sec:future} and \ref{sec:conclusion}, respectively.

\section{Background}
\label{sec:background}
\subsection{Business Process Management}
\label{sub:bpm}
\begin{figure}[tbh]
	\centering
	\includegraphics[width=\linewidth]{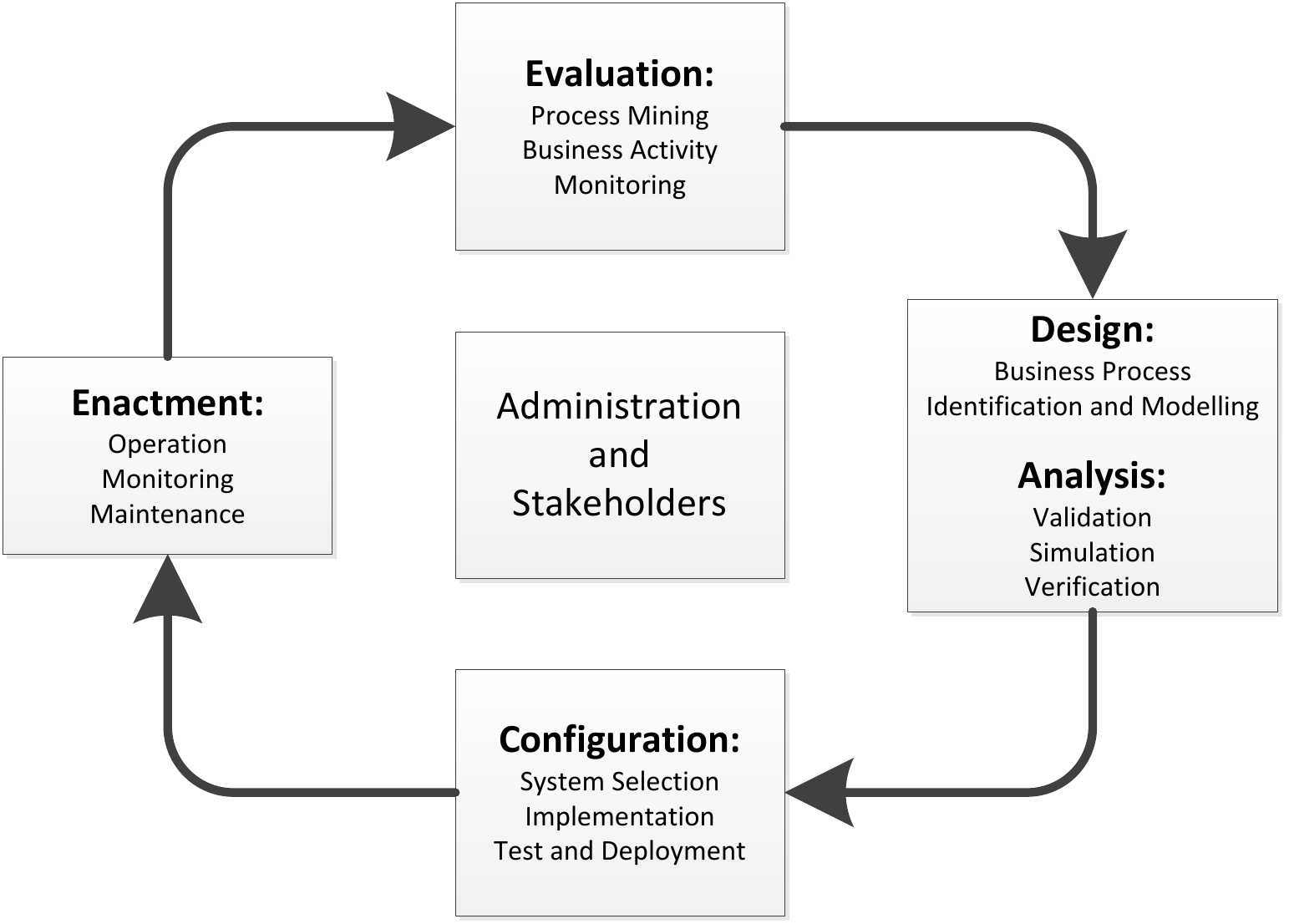}
	\caption{Business Process Lifecycle (adapted from \cite{Weske2012})}
	\label{fig:lifecycle}
\end{figure}
Processes are generally defined as sequences of tasks performed within or across companies or organizations \cite{omg11}. BPM refers to a collection of tools and methods for achieving an understanding of, managing, and improving an enterprises' process portfolio \cite{zurmuehlen10}. Within this portfolio, processes are commonly classified into value-adding core processes and non-value-adding supplementary processes. Whereas core processes are considered to contain corporate expertise and produce products or services that are delivered to customers \cite{Davenport1993}, supplementary processes facilitate the ongoing operation of the core processes. For instance, processes concerning design and production are usually seen as core, human resources processes as supplementary.
As opposed to business process reengineering, which is a greenfield methodology, BPM considers planning, controlling, and monitoring of intra- and inter-organizational processes with regards to existing operational sequences and structures in a consistent, continuous, and iterative way of process improvement \cite{BeckerKugelerRosemann2011}. As can be seen in Figure~\ref{fig:lifecycle}, the BPM lifecycle usually consists of the following phases: Design \& Analysis, Configuration, Enactment, and Evaluation \cite{Weske2012}.

The technical means to support BPM are realized through BPMSs, allowing to store process definitions (i.e., process \textit{models}), manage the enactment of process instances and monitor them, and perform logging. Process models specify process \textit{orchestrations} which comprise activities along with execution constraints and relations between them. A model is a formal description of a business process where single atomic work units (\emph{tasks}) are assigned to agents for execution. A task can either refer to work which needs to be performed manually, or to work which can be done automatically by a system. Following the principle of service-oriented computing, applications fulfilling automated tasks are commonly implemented as services \cite{VanDerAalst2003}.

\begin{figure*}[t]
	\centering
	\includegraphics[width=\linewidth]{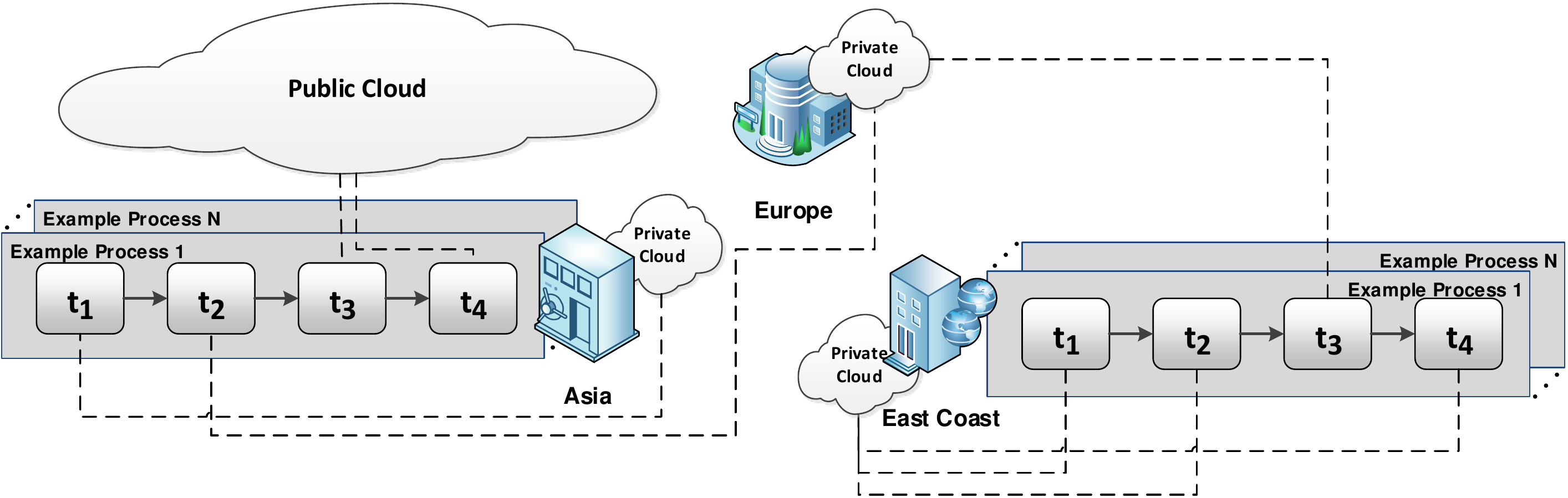}
	\caption{Example Scenario}
	\label{fig:example}
\end{figure*}

The automation of a sequence of tasks within a process by a system is often also called a \textit{workflow} \cite{ludaescher09}. An instantiated process model which is currently being executed is called a process \textit{instance}. A BPMS executes and controls these process instances. Process requesters and service providers may define QoS constraints for process models, instances or single tasks in Service Level Agreements (SLAs): execution deadlines (i.e., a maximum turnaround time for the process or task) and cost are the metrics most commonly regarded.

Process \textit{choreographies} between multiple participants ensure interoperability between process orchestrations of different participants in a collaborative process landscape. A process \textit{landscape} is made up from a large number of concurrent processes. The process landscape may span different organizations \cite{breu13}. To allow for such a collaboration, the BPMSs executing process orchestrations communicate with each other by sending and receiving messages \cite{Weske2012}. It is important to acknowledge that any process landscape is to some degree hard to predict and ever-changing, since requests to execute process instances (process \emph{requests}) may arrive at any time. 

\subsection{Cloud Computing \& Elasticity}
\label{sub:cloud}

Cloud computing enables protected access to a shared pool of configurable computing, storage, or networking resources as well as applications, which can be tailored to the consumer's needs. Cloud resources can be rapidly provisioned and released, and are billed based on actual use -- thus reducing initial investment cost \cite{Armbrust10,Mell2011}. The cloud computing paradigm  does not make other software or resource provisioning paradigms obsolete but outlines how infrastructures, platforms, and software can be commoditized and made available as a service rather than as bespoke or as a product. The approach also makes otherwise expensive resources attractive for small and medium-sized enterprises (SMEs). 

Cloud computing usually distinguishes Software as a Service (SaaS), Platform as a Service (PaaS), and IaaS \cite{Mell2011} in different deployment models: While SaaS is an approach for making software ubiquitously available as a service, PaaS provides development and deployment components to architects or developers as a service so that they can engineer applications based on these resources. IaaS enables users to access virtualized resources via a network. This can include storage, servers or network components and virtualizes physical infrastructure. 

While instant remote access to configurable services is a key aspect of cloud computing and often forms the baseline argumentation of cloud offerings, it is only the tip of the iceberg. Providing practically infinitive resources in an adequate timeframe is a feature less visible but highly important when talking about BPM in the cloud. To cope with varying workloads, cloud infrastructure services feature \textit{elasticity}, i.e., on-demand provisioning of additional infrastructure resources which are perceived to be available in unlimited numbers. Cloud elasticity is more than just resource elasticity, i.e., the ability to scale computing, storage and/or network capacity. Scaling can be done either horizontally (\textit{in/out}), i.e., additional VMs are (re)leased if necessary, or vertically (\textit{up/down}), i.e., the size of a particular VM is changed \cite{caceres10}. 

Copil et al.~\cite{copil13} additionally contribute the notions of cost and quality elasticity. Cost elasticity refers to the ability to obtain the same service at different price points through different market mechanisms, such as spot vs. commodity markets. Quality elasticity refers to the ability to trade-off QoS against cost by using cheaper services, such as VMs with fewer resources. Elasticity requirements can arise from the application, component, or programming level \cite{copil13}.

Business processes realized on top of such an elastic infrastructure have been termed \emph{elastic processes}~\cite{dustdar11}. Elastic processes go beyond a mere change to the underlying computing infrastructure. When realized, elastic processes would be able to use the infrastructure to flexibly adapt to dynamic changes in the environment, requirements, and data. It is hence necessary to facilitate self-adaptation of BPMSs with regard to leased/released cloud-based computational resources. In the following, we focus on this aspect of cloud computing rather than the mere accessibility of a ``BPMS in the cloud''.

\subsection{Example Scenario}
\label{sub:scenario}
In order to provide a basic system model for elastic processes, we describe a simplified example scenario from the financial industry. This example is illustrative only, since elastic BPM can be a solution in every domain which features extensive process landscapes that are hard to predict, including a large number of changing processes instances and/or the need to process large amounts of data in particular work tasks, e.g., eHealth \cite{mans10}, manufacturing \cite{schulte14}, or Smart Grids \cite{rohjans12}.

15\% to 20\% of banks' overall administrative expenses are attributed to the IT cost \cite{lampe13}. Even when considering only a lower bound of 1\% of the IT budget being attributed to computational resources, this is still a major expense post in the banking industry. Since the need for computational resources in business processes may vary throughout the day, using cloud-based computational resources is an obvious approach. In fact, elasticity and scalability have been named as important motivators for cloud adoption in the financial industry \cite{gill11}. Also, business process enactment using cloud resources has been discussed, however without taking elasticity into account \cite{rabhi12,shi10}. 

Figure~\ref{fig:example} depicts a simplified business process scenario from the banking industry. It concerns an international bank with branches in Asia, Europe and at the U.S. East Coast. Each of these areas operates its own data processing facilities, which provide capabilities of a private cloud, i.e., apply virtualization and scalability. Processing power is not only needed for long-running data analytic processes, but also to carry out shorter trading processes, risk assessments, or credit approval processes. Since the amount of data and the number of process instances that need to be handled concurrently may vary to a very large extent, it is difficult to predict the required computational resources at each site. Therefore, the data processing facilities are shared between the data centers, if necessary. Also, offices may use resources of other sites to avoid that very large amounts of data have to be transmitted (``code to data''). In addition, public cloud resources can be leased to extend the available computational resources, if the bank's private cloud capacities are fully utilized. However, due to regulations and compliance requirements, not all banking processes are qualified to be outsourced to a public cloud \cite{lampe13}.

Figure~\ref{fig:example} of course shows only a very small and simplified excerpt from the bank's process landscape. Most importantly, the bank is involved in a large process landscape which includes different business partners, and processes are a lot more complex than depicted. Furthermore, the figure does not capture the dynamics of the process landscape -- new process requests may arrive, processes may fail or be changed, and processes are successfully completed. 
Nevertheless, the example highlights the most important entities in elastic processes: 
\begin{enumerate}
	\item A (distributed) \emph{process landscape} is made up from a large number of process models and process instances. The process landscape is volatile and may change rapidly. 
  \item \emph{Service composition} is applied to enact the process instances.
  \item \emph{Computational resources} are provided by the cloud; both private and public cloud services can be exploited.
\end{enumerate}
\section{Elastic BPM High Level Architecture}
\label{sec:architecture}

Elastic processes require nimble, agile BPMSs able to deal with the inherent dynamism of the cloud environment and process landscape. This motivates rethinking the traditional BPMS architecture to create an elastic BPMS (eBPMS). In this section, we discuss a high level eBPMS architecture and discuss its core components. It has to be mentioned at this stage that elastic BPM is not merely a purely technological problem with a technological solution. BPM commonly involves manual tasks which are performed by human beings. While we propose to use the elasticity of cloud computing to improve BPM technology, it does not solve the issue of delays and process terminations created by a lack of personnel to attend to open worklist items. These managerial elasticity aspects have to be taken into account when implementing such a system in an organization. As for this paper, we focus on the technological challenges in realizing the proposed architecture. These infrastructural challenges form the basis for the rest of the paper.

\subsection{An Architecture for an Elastic BPMS}
\label{sub:metamodel}

\begin{figure}[thb]%
\includegraphics[width=.96\columnwidth]{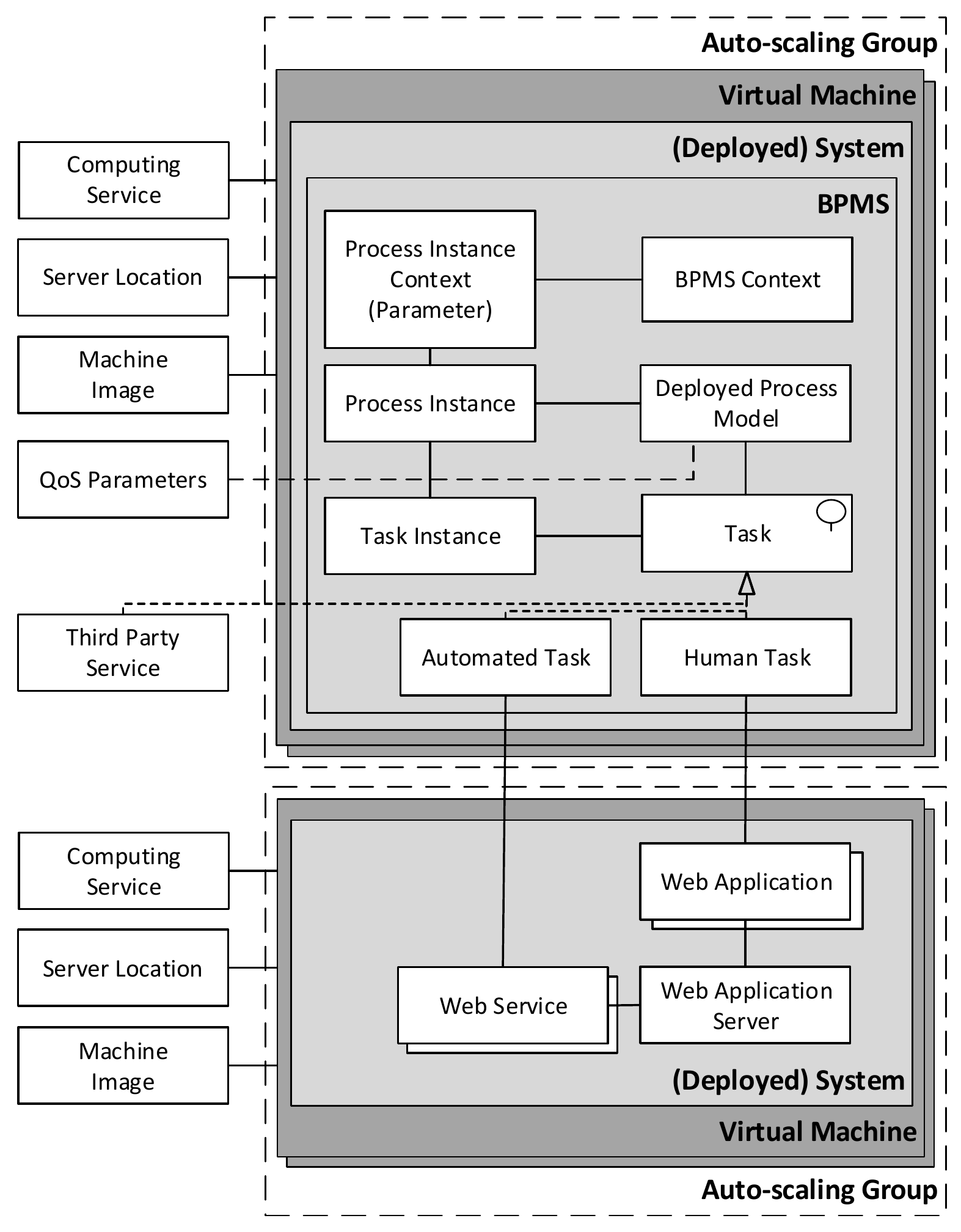}
\caption{Deployment Architecture Meta-Model for eBPMS}
\label{fig:meta-model}
\end{figure}

We map out the architecture of an eBPMS in form of a deployment architecture meta-model, describing which object types are of relevance, and which relations exist among them. Figure~\ref{fig:meta-model} shows an overview of this meta-model. Our description is focused around the core components needed for an eBPMS and hence remains at an introductory level -- a more formal specification of the meta-model can be found in \cite{janiesch14}.

As can be seen in the figure, the central entity surrounding an eBPMS setup is a VM. It is based on a VM image (or template), has a physical location, and is provided by a computing service (such as Amazon Web Services -- AWS). 
The VM is also the main component enabling elasticity: when required, more instances can be created from the same VM image, hence serving the same purpose; when less VMs are required, some of the VMs performing the same function can be terminated, although this needs to be done with care and consideration for the state of the contained applications -- see Section~\ref{sub:statemanagement}.

Inside a VM, several systems can be deployed -- where Figure~\ref{fig:meta-model} focuses on the systems relevant to eBPMS. A system can contain other systems. For instance, a Web Application Server (WAS) can contain a Web Service (WS) container, which can contain various WSs, or a WAS can contain a BPMS, or a BPMS can be stand-alone. 

A BPMS is the key component in our architecture. This BPMS has a context and a collection of deployed process models.  
The BPMS is \emph{online}, that is, process requests are continuously being received and process instances are continuously launched during its lifetime, as discussed in Section~\ref{sub:scenario}.  

A BPMS for elastic processes needs the capabilities to support the complete process lifecycle as described in Section~\ref{sub:bpm}. This includes the functionalities to request process instances, schedule process steps, lease, release and allocate cloud-based computational resources, and monitor the process enactment in order to be able to react to faults and other unplanned events. If comparing traditional BPMS and eBPMS, the major extensions and most significant modifications are related to the actual process enactment, e.g., process instantiation, process activation, monitoring facilities, tracking the current and future system landscape, reasoning about optimally utilizing resources under given QoS constraints, and controlling necessary actions (e.g., start/stop servers, move services).

We distinguish human and automated tasks, and tasks implemented by third-party services. The core distinction is that we cannot control many aspects of third party services, but we can control the infrastructure underneath services implementing automated and human tasks executed within our sphere of control. In particular, we can scale up/down/in/out the resources available to a particular task, and we can control the pipeline and achieve fine-granular scheduling.

While Figure~\ref{fig:meta-model} shows all eBPMS entities in a single VM, most of the parts are optional and may or may not be deployed within one VM or can be multi-instantiated. In Section~\ref{sub:autoscaler}, we will describe a concrete eBPMS implementation with its constituent components deployed in different VMs. The only constraint on the performance of an eBPMS is the amount of resources available in the VM(s) hosting it.

An eBPMS can create process instances from process models, and execute these instances according to QoS constraints associated with the models. Tasks in the model are instantiated for process instances as well, and each process instance has its own contextual parameters. 
Since there can be multiple instances of an eBPMS deployed in a cloud infrastructure, there is a potential that different process instances may invoke the same services separately, concurrently, and repeatedly. 

Finally, VMs performing the same function may be grouped into Auto-Scaling Groups (ASGs), which control the number of functionally equivalent VMs. Typically, the identity of a VM in an ASG is not of importance -- only that there are $x$ VMs, all instantiations of a certain machine image, present in the group. The parameter $x$ can be controlled manually, via an API (i.e., a custom scaling controller), or by cloud provider-specific scaling mechanisms. To spread the load over the available VMs, load balancers can be used. These can be independent of ASGs, but cloud providers like AWS offer both services in an integrated fashion, where new instances in an ASG are automatically registered with the respective load balancer (called Elastic Load Balancer -- ELB), and retired instances are deregistered automatically. It is important to be aware of the VMs' states as they might be crucial for the successful completion of a process. For example, despite being idle, a VM which is still waiting for a process reply cannot be terminated without compensation on the process side.

\subsection{Challenges for Elastic BPM}
\label{sub:elasticbpm}

The deployment architecture meta-model presented above describes a dynamic environment in which cloud-based computational resources and the eBPMS are subject to varying workloads. As mentioned before, elasticity of cloud infrastructure allows systems to meet dynamic resource demands. 

Existing BPMSs could be adapted to the proposed deployment meta-model. However, this would ignore the many issues and challenges that would become apparent when the resulting eBPMS is used in production. The first issue would be the scheduling of tasks in process instances. Resource elasticity makes it possible to spawn enough VMs to allocate all tasks at once. This however ignores the interdependencies between tasks in a process instance as well as the exorbitant cost that this would entail. However, elasticity presents new opportunities for optimizing task schedules. For instance, cost elasticity, such as the notion of spot instances, can be employed to speed up process enactment through selective parallelization without much additional cost. Therefore, elasticity introduces a new perspective from which to tackle scheduling for process instances.

This leads us to the allocation of resources to tasks. While in theory, resource elasticity is assumed to be instant, in practice, there is a delay between provisioning a resource and the time when it is available. This motivates the need for methods to predict demand so as to provision resources in advance. QoS constraints defined for process instances can also influence resource allocation. For example, strict QoS requirements may lead to overprovisioning, where more resources than needed are allocated so that QoS requirements are always met. 

Effective prediction methods hinge on extrapolations from detailed performance data obtained from previous process instantiations, task executions, service invocations and provisioning operations. Recent developments have led to a deluge of performance data that needs to be processed as fast and accurately as possible in order to feed into resource planning. This motivates rethinking of process monitoring in BPMS.

The discussion so far has assumed a central decision-making component, a \emph{controller}, that keeps track of the global state of a BPMS including status of all process and task executions, and the resource conditions. However, with the increase in the scale of a system, the number of variables under observation increases exponentially. The central controller becomes a bottleneck for the BPMS. A decentralized architecture enables the controller to be decoupled and its functions distributed among a number of nodes in the system. However, decentralization brings with it problems of coordination as well. 

Last but not the least, current BPMS largely assume that the state of the system is knowable, is reflected in a database and is consistent throughout the duration of process enactment. However, resource elasticity induces huge variations in the configuration and state of an eBPMS and its context. Thus, the knowledge of the eBPMS may be out of sync with the true state of the system. This could cause problems with task execution. 

We posit that realizing a truly elastic BPMS involves meeting each of these infrastructural challenges, namely \emph{scheduling}, \emph{resource allocation}, \emph{process monitoring}, \emph{decentralized coordination}, and \emph{state management}. In the following section, we will discuss the current state of the art in these areas.
\section{Infrastructural Challenges for Elastic Processes}
\label{sec:challenges}
\subsection{Scheduling for Elastic Processes}
\label{sub:processcontrol}
(Business) Process scheduling is defined as the problem to find an execution sequence for the process so that its tasks are finished under given constraints, e.g., timing, causality, and resource constraints \cite{eder99,avanes08}. Especially scheduling under causality and timing constraints has been a major research topic in recent years, while resource constraints are a relatively new topic \cite{huang11}. Scheduling under resource constraints describes the function of analyzing a queue of process instances or tasks, respectively, deriving the demand for (computational) resources, and allocating tasks to these resources \cite{yu08a}. For elastic processes as regarded within this paper, these distributed computational resources are provided through a public or private cloud or a mixture of both. The basic goal of process scheduling is the \textit{schedule}, i.e., a description of which tasks will be executed at what point of time using which computational resources. Instead of just meeting these constraints, scheduling usually aims at maximizing QoS subject to the given constraints, e.g., through defining an optimization problem. Since the execution time of a particular task is only partially known in advance, scheduling for elastic processes is a \textit{stochastic} scheduling problem \cite{baker09}.

In an elastic BPM scenario, the process queue can change rapidly due to newly arriving process requests or since computational resources do not perform as expected. Hence, it might be necessary to quickly adapt the schedule. This poses a problem, since optimization for process tasks (here: provided through cloud-based computational resources) under given constraints is a NP-hard problem \cite{strunk10} and there is no known way to decide such problems in polynomial time \cite{garey79}. Hence, heuristics and approximations need to be found to compute a schedule. A very important input for process scheduling is knowledge about the runtime of tasks on particular resources. In most existing work, this is done by defining such a time span. However, for cloud resources and elastic processes, this information needs to be derived from historical monitoring data (see Section~\ref{sub:monitoring}), where some degree of uncertainty is unavoidable~\cite{Schad10}.

In this subsection, we discuss research on scheduling for elastic processes. First, we briefly discuss approaches in traditional BPMS, then, scheduling for SWFs and grid workflows, and finally elastic processes are addressed.  

In traditional BPMS, jobs are processed rather flow- than schedule-oriented, i.e., tasks are attributed to human- or machine-based resources and these resources pick their next task themselves without taking care of deadlines \cite{mans10}. The theoretical foundations for process scheduling can be traced to the fields of Operation Management and Operations Research, where \textit{jobs} are scheduled on a number of machines \cite{baker09,baggio04}. In particular, scheduling for (flexible) flow shops and job shops \cite{pinedo12} is related to the field of elastic processes, since basic assumptions are the same for such shops and elastic process enactment. However, in their basic forms, neither flow shops nor job shops take into account scalability of the resources.

Within traditional BPMS, which usually address both human- and machine-provided tasks, scheduling is mostly considered with regard to timing and control flow, e.g., \cite{eder99}, or instead of scheduling based on deadlines, a first-in-first-out (FIFO) approach is applied \cite{baggio04}. While there are some approaches to include time constraints in BPMS, e.g., \cite{combi06,lanz12}, resource allocation is usually not regarded together with timing \cite{senkul05}. This is however a major objective for elastic processes, since resource utilization and scheduling have an effect on each other. Especially the cost constraint is directly affected by the chosen resource allocation. 

There are several approaches to scheduling for business processes, which have different optimization goals like reducing the number of late processes or the mean tardiness percentage \cite{baggio04}. However, none of them can be directly applied to elastic processes, since a fixed amount of resources is foreseen \cite{baggio04, combi06}, cost are not regarded \cite{baggio04}, and/or only single process instances are taken into account \cite{combi06,senkul05}.

Several approaches for workflow scheduling for grid computing have been analyzed in a survey by Yu and Buyya, with a specific focus on SWFs \cite{yu08a}. Since this survey provides a very comprehensive overview of such scheduling approaches, we abstain from another in-depth discussion, but want to highlight that scheduling algorithms (for SWFs) in grids are only partially applicable to elastic processes, especially since the aspect of resource allocation (see Section~\ref{sub:allocation}) in terms of VM provisioning is not taken into account \cite{frincu13}.

Scheduling for SWFs has also been proposed with regard to cloud-based computational resources and cost minimization, e.g., \cite{pandey10}. Some approaches take into account overall runtime of single SWFs \cite{szabo12} or even user-defined deadlines \cite{byun11, abrishami13}, but concurrent workflows are not regarded. 

While the discussed approaches from the field of SWF scheduling offer interesting ideas and insights, there are certain differences between business processes and SWFs which prevent a direct adaptation of such approaches \cite{ludaescher09}. First, SWFs and business processes differ in their goals (``experimental'' vs. ``business-driven'') and therefore, the QoS requirements of the latter are of much more importance. Second, the number and characteristics of process instances are different, as for business processes, it is common that a large number of identical processes is carried out concurrently, while for SWFs, many interdependent, (slightly) different instances of the same workflow are carried out in batches. Third, SWFs are usually dataflow-oriented, i.e., the execution flow follows the dataflow; this explains the focus on data exchange applied by most of the approaches. In contrast, in business processes, the control flow is explicitly modeled while data-related interdependencies between process instances are only of secondary importance \cite{vdAalst03a}. As a result, existing approaches from the SWF field are not directly applicable to business processes.

The number of scheduling approaches explicitly aiming at concurrent elastic processes is still straightforward: There are scheduling heuristics aiming at reducing the mean cost and mean turnaround time of all process instances while taking into account SLAs for single instances \cite{xu09}. Other approaches disregard SLAs, and the optimization is either Pareto-optimal or based on weights or priorities indicating the importance of cost and turnaround time \cite{juhnke11,bessai13a}. Recent investigations have considered SLAs in terms of deadlines for single tasks and single process instances, but only sequential processes are supported \cite{hoenisch13,hoenisch13a,schulte13a}. Notably, all these approaches to scheduling for elastic processes also take into account resource allocation. 

It should also be noted that there are several approaches to schedule single QoS-constrained applications on cloud-based computational resources, e.g., \cite{leitner12,wu11}, partially even taking into account that applications are made up from parallel tasks, e.g., \cite{vdbossche13}. These approaches all aim at cost-efficiency and do not take into account the process perspective.

\subsection{Resource Allocation}
\label{sub:allocation}
As has already been discussed in the section on process scheduling, BPMS traditionally use a fixed (and thus limited) amount of computational resources for executing processes, if resources are regarded at all. With the advent of cloud computing, however, BPMS can benefit from the full potential of scalable cloud resources and can dynamically use as many resources as actually required. Processes with varying demands of computational resources can, thus, be executed on virtualized infrastructure to scale on-demand. 

Business processes orchestrate tasks as service requests in a structured, procedural fashion. This process knowledge is valuable additional data for predicting resource demand, which current infrastructure scaling mechanisms do not take into account. Hence, this pre-existing knowledge present in (e)BPMS can and should be used to improve infrastructure resource controllers through the process-based alignment of available computational resources with the respective demand. Otherwise, the behavior of business processes with long-running transactions may cause bottlenecks which currently cannot be handled. For example, a process may at one stage build up too large a backlog which is processed simultaneously through a scaled-up ASG. Any independent, subsequent auto-scaling services will be overburdened, since they were not prepared to receive such a significant amount of process instances at once.

This motivates the development of a process knowledge-aware IaaS resource controller instead of the non-BPM-aware auto-scaling services commonly available in commercial clouds. As we will discuss shortly, present researchers have also not provided enough consideration to the problem of using process knowledge for auto-scaling. 

While scheduling deals with distributing workload over time~slots of available resources, resource planning is concerned with \emph{predicting} future requests and workload, \emph{pro-actively planning} how many resources of which type to \emph{acquire/release}, and doing so when needed. A design alternative to prediction and planning is to have a purely \emph{reactive} system, i.e., to base decisions on acquiring/releasing resources solely on current (or recent) monitoring data. Of course, hybrid solutions are possible as well: predict, plan, but if reality deviates from the prediction, react quickly. 

For predicting request arrival frequency, traditional techniques such as predicting values from time series data can be used, e.g., \cite{Dietterich02,Bermolen09}. Such techniques can work in complete autonomy, and make predictions on the request frequency purely by observing incoming requests to date. 
In contrast, there are event-aware techniques such as \cite{Sladescu12}, where the operator of a system can manually inform the prediction engine about planned events -- such as start of ticket sales for big events, marketing promotions, product announcements, etc. The expectation is that the request profile during an event may look dramatically different from the usual request profile.

In order to estimate upcoming workload, it is necessary to predict the number of requests made to a process model, and how much workload a request causes where. In the context of elastic processes, requests are made with regard to a particular process model. These can be requests triggering new instances of a model, or intermediate message exchanges. As for intermediate messages, these can either influence workload, e.g., by triggering a stream of the process that has been waiting for this message, or not. In the former case, the workload caused by the message can be regarded similarly as the workload triggered by a request that creates a new process instance; in the latter case, the request does not influence the workload. There have been a number of publications discussing workload prediction for traditional BPMS~\cite{Marzolla07,Reijers05,Heinis05}.

In order to serve the predicted workload, it is necessary to acquire resources in a timely fashion. For instance, if it takes 3 minutes from requesting a new VM until it is available, booted up, and all application logic prepared for serving requests, then this lag needs to be planned for. 

Typically, public cloud providers offer resource types at certain levels of service, e.g., a VM with 4 cores, 16GB of RAM, etc. The selection problem is thus a discrete one: how many resources of which size shall be requested? In contrast, private cloud and virtualization techniques can offer much more fine-grained vertical scaling, e.g., by allowing to specify the amount of RAM in small increments.\footnote{For instance, VMware vCenter Server 5.5 (\url{https://www.vmware.com/support/pubs/vsphere-esxi-vcenter-server-pubs.html}) allows to set the amount of RAM in increments of 4MB, from a minimum of 4MB to a maximum of 1011GB, giving theoretically 258,815 distinct options.} The question then is how to split the available physical hardware into chunks of virtualized resources, and the finer steps of course mean that there are a lot more options to consider.

Finally, it is necessary to have an estimation model of the capacity of resource types -- from load balancers over the network capacity to computing power, RAM, disk speed, etc.
Given such a capacity model, it is possible to estimate how the relative load on each resource metric will develop, based on the predicted workload.

An alternative to request prediction is to focus purely on the current load of resources, and make scaling decisions reactively. At any rate, resources must be acquired and released to accommodate changes in demand. In the general case -- i.e., without regard to the specific needs of elastic processes --  acquisition of cloud-based computational resources for single services and applications has been investigated by a number of researchers. Some have focused on cost-efficient resource allocation, e.g., \cite{lampe11}, and the enforcement of SLAs, e.g., \cite{buyya10,Cardellini2011}. There are also approaches which aim at achieving a higher resource utilization \cite{juhnke11,emeakaroha_managing_2013,li11}. Usually, resource acquisition is done in a reactive way based on rules. Other researchers have used machine learning to automatically scale an application up or down \cite{li11}. All these approaches do not take into account BPM, but instead focus on reactive scaling for single applications. 

Furthermore, scaling in any way can require follow-on configuration. For instance, once a new VM is available, it needs to be registered -- e.g., in an ASG -- in order for it to receive traffic. Also, the network configuration might need to be updated. Removing a VM needs to be done in the opposite order: deregistering it first, waiting until all running operations have been completed, and then shutting it down. 

Depending on the cost models of the given cloud provider, it may not make sense to shut down a VM at an arbitrary point of time. For instance, AWS charges VM usage by the hour. Thus, rather than shutting down a machine that is no longer required, one may rather keep it in a pool of VMs on standby until a few minutes before the next billing time unit starts. Details of such a strategy can be found in \cite{euting14}. 

Resource acquisition with regard to elastic processes has been discussed as a ``side-effect'' of process scheduling by a number of researchers (see Section~\ref{sub:processcontrol}). Apart from that, there are very few approaches, which in any case have some limitations, e.g., since it is not allowed that service instances running on the same VM may be shared between concurrent processes \cite{wei13}. As a result, the potential for optimization of resource allocation is not completely exploited. Related to approaches directly aiming at elastic processes, there are approaches for multi-service web applications, with the different applications being composed in a directed acyclic graph \cite{dejun10}. It is unclear as yet if such techniques could be adapted to BPM scenarios with potentially complex control flows.

\subsection{Process Monitoring and Data Collection}
\label{sub:monitoring}
The concept of Event-Driven Architectures (EDAs) is a recent addition to the list of architecture paradigms for system landscapes. It often complements service-oriented architectures (SOAs). It emphasizes the orchestration of applications and processes through events which can originate anywhere from external sensors to internal or external business IT systems \cite{Luckham2002}. The core idea is to enable systems to anticipate and react to normal and abnormal event types and adapt the execution of underlying business processes.

An event is defined as a data object that is a record of an activity that has happened, or is thought of happening in a system \cite{Luckham2002}. The event signifies the activity. Such objects comprise, among others, activities, actors, data elements, applications, or entire processes. Due to the high volume, velocity and possibly variety of simple events, events may need to be correlated and aggregated to complex events to provide meaningful insights. This requires scalable processing technology. Complex event processing (CEP) is such a technology. It comprises a set of techniques for making sense of the behavior of a system by deriving higher-level knowledge from lower-level system events in an online fashion by filtering, pattern matching, and transforming \cite{Etzion2010}. Through the definition of abstraction relationships, it is possible to define complex events by aggregating a set of lower-level events. Hence, a complex event signifies a complex activity which consists of all the activities that the aggregation of individual events signified. Orthogonal to the concept of aggregation is the concept of causality. Lower-level events might happen in dependence of each other. Causality plays an important role in any kind of root cause analysis, both online and ex-post.

Recently, event-based or event-driven BPM approaches have emerged \cite{Janiesch2012}. In the context of eBPMS, the ability to define business rules dynamically and correlate events from a variety of resources is crucial. Events in BPM can root from distributed IT systems which are not accessible by consumers and potentially federated amongst multiple entities. Here, CEP emerges as a valuable instrument for a timely analysis of BPM- and other service- or infrastructure-related events. The interaction of CEP and elastic BPM can be bi-directional, as other systems can act as (a) a producer and as (b) a consumer of events \cite{Etzion2010}.

CEP as part of elastic BPM can be applied to monitor and control elastic processes in real-time. This extends upon the real-time concept of business activity monitoring by leveraging the capabilities of CEP to include decision making and feedback mechanisms to the originating service network and other systems. This effectively means that a CEP engine, in parts, manages rather than observes the execution of the service networks. Hence, one could also speak of business activity management rather than of monitoring \cite{Janiesch2012}.

Any architecture for elastic event-driven BPM is based on the typical tri-partition in CEP. In most CEP applications, there is a distinction between the entities that input events into the actual CEP system (event producers), the entities that process the events (event processors), and the entities that receive the events after processing (event consumers).

The major event producers in BPM are individual systems (application servers) providing orchestrated or atomic services. In cloud-based process choreographies, process enactment may be distributed amongst independent service providers having varying levels of autonomy and potentially a multitude of different systems. Typically, every eBPMS provides an environment for the execution and interaction of services, based on existing process choreography definitions. These systems' logging components offer functions to track the progress of execution and log corresponding audit data including data on occurred state transitions of choreography-relevant objects (e.g., processes, services, resources, work items). This audit data constitutes a rich source of service execution events for subsequent analysis, but is usually not accessible from outside the individual execution engine. Hence, an event publisher \textit{output adapter} is required to make the data available for participating parties in a service network. The main functions of this output adapter is \textit{sniffing}, \textit{formatting}, and \textit{sending} of events \cite{Luckham2002}. 

In the sniffing step, the adapter detects events in the local audit data (service orchestration) and other relevant sources (e.g., work item lists or cloud-based computational resources). Typically, the sniffer maintains a list of filters that defines which events should be observed and subsequently extracted to prevent sensible data from being published or to simply reduce the volume of events. Such filters can be defined on type level or instance level. It is important that event sniffing must be benign, i.e., it must not change the service network's behavior. After detecting an event, the corresponding data has to be extracted and transformed to a common event format.

However, not all system events can be easily mapped into a standard format. Efficient or even standardized monitoring of processes or cloud resources is still in its infancy. As of now, only basic health information on cloud resources is made available for service consumers. Amazon's CloudWatch API, for example, allows for the monitoring of basic information like minimum, maximum or average CPU load, memory usage or request latency through a standardized API. Correlation of resource events to process events (or vice-versa) is not supported as of now, as existing cloud resource management services do not incorporate the notion of processes or service choreographies. Through the application of this concept of event output adapters, existing cloud resources can be integrated into an  event-driven eBPMS allowing for an aggregation of events towards higher-level network events like potential performance slowdown as system metrics reach thresholds.

Besides eBPMS and cloud infrastructure, there are a number of other systems that can act as event sources: sensors, for instance, can produce events that signify the actual execution of a physical service activity or report on real-world events that are happening in the context of the execution (e.g., weather changes, traffic conditions, position and movement of objects). Transactional systems, such as ERP or CRM systems, can produce events that are related to choreographies but lie outside of the control sphere of a service network (e.g., low stock alert, creation of a master data record).

The raw events sent out by the various event producers are received and processed by an event processor, i.e., the actual CEP system. Positioning CEP as a central component in this event-driven measurement and management architecture is crucial. CEP systems are typically deployed in a centralized fashion, however, there are early versions of CEP engines which can be hosted in the cloud or in a distributed fashion. In any case, queries can be staged and distributed so that the risk of a single point of failure within the service network can be minimized. Through virtualization, placement of the event processor (in any of its occurrences illustrated below) in the cloud can reduce its susceptibility. Events are processed in an event processing network as introduced at the beginning of this section.

After processing, events are queued into an output event stream, ready to be forwarded to possible event consumers. The outgoing data does not necessarily have to be represented as an event anymore; it can, for example, be in the form of Key Performance Indicators (KPIs), messages, or remote function calls.

Finally, an event consumer is an entity that receives and visualizes or acts upon events from a CEP system. It is the logical counterpart of the event producer. Dashboards, messaging services, and choreography participants are the main event consumers in an elastic BPM scenario. An event output adapter of a CEP system might, for instance, initiate new elastic process instances or lease or release virtualized infrastructure. However, as of today, most BPMS can only be called via SOAP or REST messages. Likewise, manipulating or updating the conditions of a choreography execution is next to impossible today, although corresponding resource allocation might have a major impact and change decisions which control the flow of a choreography instance \cite{becker12}. An API call to allocate more resources for a particular service (e.g., in case of dropped availability or responsiveness) or suspend, resume, or cancel running service instances (which could be anything from a single service to a choreography) could lead to an increase in needed computational resources. A comprehensive summary of the current state of the art of event-driven BPM can be found in \cite{krumeich14}.

\subsection{Decentralized Coordination for Process Enactment}
\label{sub:decentralised}

Traditionally, a BPMS has been a single component that parses the process model to create process instances and tasks, schedules the tasks onto available resources and monitors their execution.  A centralized BPMS exhibits poor scalability as it becomes a single point of failure and congestion in the network~\cite{alonso_exotica/fmqm:_1995}. It would also not be scalable while handling large numbers of long-running, data-intensive process instances, as it is the case in an elastic process landscape. Furthermore, clouds are dynamic systems with multiple variables such as workload and resource conditions changing simultaneously. A centralized component may therefore not be able to quickly react to changing conditions, especially when large numbers of resources are involved.  Therefore, these are not able to take full advantage of rapid elasticity of cloud platforms. 
 
However, in recent years, with the advent of grid and cloud computing, there has been a rethinking on the architecture of the BPMS itself. This has led to extensive research on decentralized process management in recent years~\cite{alonso_exotica/fmqm:_1995,taylor_triana_2007, rahman_cooperative_2010, legrand_workflow_2011}. One model of decentralization involves a master enactment engine that delegates responsibility to slave engines for managing a subset of the process tasks or functions such as data management~\cite{duan_dee:_2005}. Another model involves independent components that manage disjoint workflow executions yet coordinate task execution and resource management with each other via a peer-to-peer (P2P) network. 
 
Many projects have employed P2P networks as the communication substrate. SwinDeW-C~\cite{liu_swindew-c:_2010} is a decentralized BPMS that uses the broadcast-based JXTA P2P protocol for communication. Rahman et al.~\cite{rahman_cooperative_2010} describe a system based on the notion of independent resource managers cooperating in a P2P environment to cooperatively execute tasks from different processes. The resource managers post information about the resources under their control to a \textit{d}-dimensional distributed tuple space that is maintained on top of a Distributed Hash Table (DHT). Process enactment is mediated by resource brokers who search for appropriate resources in the coordination space and make claims on the resources to execute tasks. Claims from different brokers are resolved by the managers according to their priority, the amount of resources requested and that available.

Stojnic and Schuldt~\cite{stojnic13} introduce OSIRIS-SR which is also a decentralized process execution engine wherein the execution of tasks is delegated to a P2P network of independent nodes. Information about the tasks, the nodes, and their load is maintained in a set of separate global repositories that are replicated via a publish/subscribe (pub/sub) mechanism. 
This bus can be maintained on top of a P2P network as well as other communication substrates.
The same motivations have led to the introduction of decentralized architectures for process orchestration and execution. BPELcube~\cite{pantazoglou_decentralized_2013} is a process execution framework where tasks are distributed over a P2P architecture organized as a hypercube. Within BPELcube, each peer can act as a manager for executing a single process instance and distribute tasks to nodes on the basis of Least Recently Used order. G. Li et al.~\cite{li_distributed_2010} describe NI\~{N}OS, a distributed process execution engine based on a P2P pub/sub system~\cite{cheung_load_2010}.  NI\~{N}OS features an agent-based architecture where the execution of a business process is distributed among a network of agents that collaborate via pub/sub messages. The messages from the WS invoked in the course of the execution are translated to and from the pub/sub messages by the agents. 

While a decentralized architecture brings with it a gain in reliability and scalability, this is balanced by the requirement of ensuring reliable communications between the components. Yet, Benatallah et. al~\cite{benatallah_facilitating_2005} experimentally demonstrate that P2P orchestration of execution of composite web services provides significant performance benefits over a centralized model.

However, all of the decentralized systems surveyed have been deployed in environments where the usage of peer nodes does not incur economic cost. While these handle situations such as peers joining or leaving the network and node failures, there is no need for empowering eBPMS functionalities to provision additional resources or shut down resources.  In cloud environments, where resources are leased on the basis of billing time units such as hours or minutes, resource elasticity can be leveraged to balance SLAs of the processes against the cost incurred in executing them.  Centralized management systems such as Amazon's Simple Workflow Service\footnote{\url{http://aws.amazon.com/swf/}}, Oozie~\cite{islam_oozie:_2012}, Nephele~\cite{warneke_exploiting_2011} and LoM2HiS~\cite{emeakaroha_managing_2013} are able to dynamically provision resources in order to meet the needs of process enactment. However, this brings with it the attendant problems of a centralized BPMS discussed previously.

While systems such as Grid-Federation~\cite{rahman_cooperative_2010} and BPELcube~\cite{pantazoglou_decentralized_2013} have the notion of automatic resource allocation and deallocation, they have not been deployed in elastic contexts where the most critical decision is when to take on the additional cost of provisioning a resource or the responsibility of removing a resource. Therefore, a decentralized eBPMS should involve local managers that are independently able to provision or deprovision nodes as needed to meet the SLAs of process instances as well as to keep cost in check. The local managers can cooperate to execute tasks and to share the load amongst themselves. Since there is no centralized authority to oversee executions, the system has to be self-regulating so as to avoid situations such as a glut or scarcity of resources due to over- or underprovisioning respectively. 

Decentralization is also core to the notion of EDAs described in the previous section. Traditional SOA-driven BPM depends on invocation of services needed for completing a task. Within EDAs, services publish events which are then subscribed to by process managers~\cite{muthusamy_bpm_2010, buchmann_calls_2012}. The latter react to these events through their actions which then result in further events. An EDA allows different components involved in process orchestration and execution to be decoupled, and some amount of control to be delegated to the service bus. Also, since the execution is reactive, this architecture is suitable for integrating elasticity actions, such as provisioning or deprovisioning instances, that are taken in response to events such as SLA violations. 

\subsection{State Management}
\label{sub:statemanagement}
There are different states associated with the enactment of a process instance, which are related to the process lifecycle (see Section~\ref{sub:bpm}). The \textit{configuration state} encompasses the setup of a BPMS and that of the service instances. However, this does not imply that the BPMS necessarily controls the configuration of the service instances. In a cloud computing setup, a service may be deployed on a set of VMs, and the eBPMS may merely be aware of their location. The \textit{process enactment state} includes the current status of the tasks under execution, list of tasks that are waiting to be executed and the dataflows that are in progress. This corresponds to the data collected and maintained by the process scheduler (see Section~\ref{sub:processcontrol}) during process enactment. In a centralized BPMS, this data is maintained in a persistent medium, such as a database, in a single location. Where the BPMS is decentralized, each of the independent components may maintain their own individual databases. In this case, the enactment state encompasses the state stored in databases as well as the messages sent between the components. Notwithstanding these, each of the service instances also has its own state maintained in a database. 

The process enactment may require an environment with specific communication, availability, performance, security and isolation conditions~\cite{schuster_configuration_1999}. In traditional application deployment, information about software configuration is stored in a central location and it is assumed to be accurate and appropriate for any condition. Configuration is changed in a controlled manner and the effects of changes are known to be predictable. This assumption breaks down in an elastic process landscape as components (here: services) of an application (here: elastic process) may be constantly redeployed among VMs as they are provisioned and deprovisioned. This may violate the conditions that were set down when the application was deployed. 

This problem of preserving correctness while applying changes to a business process in enactment is well-known~\cite{rinderle_correctness_2004}. Historically, this problem has been well-studied in the context of adaptive workflow management~\cite{han_taxonomy_1998}. This research has led to formal specifications for correctness of workflow executions with pre- and post-conditions that have to be satisfied while applying changes to the workflow structure~\cite{reichert_adeptflexsupporting_1998}. Changes allowed in this model are modifications to the controlflow or dataflow in a workflow in order to deal with dynamic conditions. However, as described in previous sections, elastic processes are subject to changes to infrastructure in which computing, storage and network elements as well as applications are added or removed on-demand. This introduces new challenges in configuration management to satisfy correctness of process enactment. Indeed, the notion of correctness for elastic processes will have to be defined as different scenarios could produce different outcomes, all of them valid. 

Recent configuration management tools such as CFEngine \cite{burgess_promise_2006}, Chef\footnote{\url{http://www.getchef.com/}} and Puppet\footnote{\url{http://puppetlabs.com/}} provide declarative interfaces to specify and control application deployment on clouds as well as to manage scaling of applications. Chef and Puppet are centralized systems relying on configurations derived from a single source of truth. CFEngine uses autonomous agents that voluntarily cooperate to maintain a stable configuration in the face of varying resource conditions~\cite{burgess_promise_2006}. However, these and other tools are primarily devoted to system management. 

Some of the research on autonomic process management in cloud environments~\cite{xu_url:_2012, wei13a} have explored agent-based and reinforcement learning approaches for adapting configurations in the face of dynamic workloads. However, the focus of these approaches is more on resource management during process enactment and less on other aspects of configuration management. Furthermore, these solutions are designed for an environment where the eBPMS is in control of the service deployment which may not always be the case, especially in business processes. Therefore, there is a need for continuous configuration checks so that the process executes to its specifications.

The process enactment state also has to be consistent throughout any operation to avoid faults. Scaling up the persistence layer is not as clear-cut as it is for the application layer. Data consistency and durability have to be considered while adding or removing nodes from a data management system. 

There are different ways in which persistent state can be adapted to elastic conditions~\cite{sakr_cloud-hosted_2014}. One approach involves storing it in relational database management systems (RDBMS) that are deployed onto VMs. Scaling up the database is performed by adding VMs. This is also similar to commercial services such as Amazon's Relational Database Service\footnote{\url{http://aws.amazon.com/rds/}}, Microsoft Azure SQL Database\footnote{\url{http://www.windowsazure.com/en-us/services/sql-database/}} or Google Cloud SQL\footnote{\url{https://developers.google.com/cloud-sql/}}. This approach provides a standard method, namely Structured Query Language (SQL), for interfacing with the database as well as strong transactional guarantees. This means that existing BPMS and services that use RDBMSs will be able to scale on cloud infrastructure without modification. However, such systems are limited in their elasticity as the operations required for scaling up are resource-intensive and time-consuming.

Another approach involves storing persistent data in so-called NoSQL data stores such as Apache Cassandra\footnote{\url{http://cassandra.apache.org/}} or 
MongoDB\footnote{\url{http://www.mongodb.org/}}. These commonly offer a weaker consistency model and limited support for transactions. They provide a variety of NoSQL interfaces, all of which are based on a reduced subset of SQL-like primitives. (e)BPMS would have to be reengineered in order to support these alternative databases. However, these offer better performance than RDBMS, particularly on cloud infrastructures. Also, it has been shown that such databases can be scaled in a lightweight and frictionless manner~\cite{li_efficient_2013}, and therefore, are able to use resources in a more efficient fashion.

\section{eBPMS Realisations}
\label{sec:realisations}
In Section~\ref{sec:challenges}, the current state of the art for different infrastructural challenges for elastic BPM has been discussed. In this section, we give examples in terms of two research eBPMS which represent the current state of the art, namely the \emph{Fuzzy BPM-aware Auto-Scaler} and the \emph{Vienna Platform for Elastic Processes}. These eBPMSs focus on resource allocation and process scheduling.

It should be noted that there are various approaches to bring BPMS to the cloud, i.e., as SaaS. Both research, e.g., \cite{mangler10}, and commercial solutions, e.g., by IBM\footnote{\url{http://www-03.ibm.com/software/products/en/business-process-manager-cloud}} or Bizflow\footnote{\url{http://www.bizflow.com/bpm-software/cloud-bpm}}, are available. However, to the best of our knowledge, there are no approaches where these SaaS-BPMSs are able to provide elastic processes as defined in Section~\ref{sub:cloud}. 

\subsection{Fuzzy BPM-aware Auto-Scaler}
\label{sub:autoscaler}
\begin{figure}[htb]
	\centering
	\includegraphics[width=\linewidth]{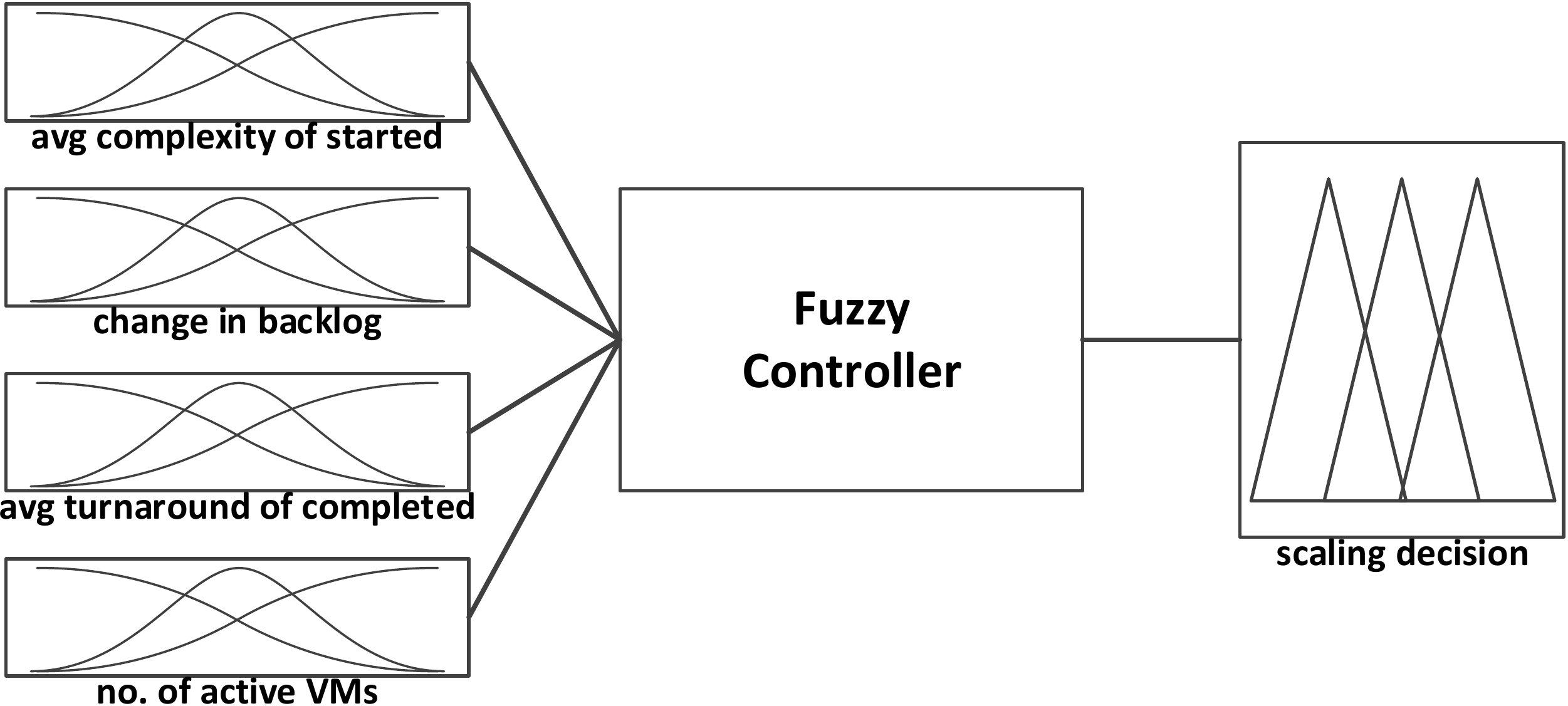}
	\caption{Overview of the Fuzzy BPM-aware Auto-Scaler}
	\label{fig:autoscaler}
\end{figure}
The Fuzzy BPM-aware Auto-Scaler is primarily a BPM-aware resource controller\footnote{For a more detailed discussion of the algorithms and implementation of the \emph{Fuzzy BPM-aware Auto-Scaler}, see \cite{euting14}.} based on a Mamdani fuzzy controller \cite{euting14,Mamdani1974}. It monitors process KPIs and VM KPIs, preprocesses the data, and evaluates it with fuzzy rules of a fuzzy control model to determine how many VMs to add or to remove, if any, from a load balancer. It then executes a scaling decision and -- after the scaling execution has completed or timed out due to failure or excessive delay -- the procedure is started again.

The scaling rules of the underlying resource controller consists of a premise, the input from one or more sensors, and a conclusion, the output related to one or more actuators. A rule's conclusion is applicable only if the situation described by the rule's premise is observed. The control rules can only be useful if the occurring linguistic terms are defined. Therefore, all appearing terms have to be defined for the respective co-domain by an associated fuzzy set. Both premise and conclusion are defined using linguistic variables and linguistic values. In fuzzy theory, linguistic terms are mathematically represented in form of membership functions describing fuzzy sets. Hence, each linguistic term of each linguistic variable has a respective fuzzy set associated. 

To derive a control action from sensor input, the fuzzy controller's control logic has to evaluate the rule base. This is done in two steps: (a) evaluate all rules individually and (b) aggregate individual results of the rules into a global result. The Fuzzy BPM-aware Auto-Scaler currently bases its scaling decisions on four input parameters obtained from the eBPMS and the load balancers of the ASGs: \emph{average complexity of started instances}, \emph{average turnaround time of completed instances,} \emph{change in process backlog}, and \emph{number of active VMs} (see Figure~\ref{fig:autoscaler}).

Average complexity of started instances provides an indicator of the dimension of the requested workload during the current control cycle. It is a value calculated from variables either populated during process enactment or used for process instantiation. The value of this variable can be used to estimate (future) resource requirements. The average turnaround time of completed instances provides an indicator of the timeliness of process instances that terminated (successfully). This provides insight into whether the preceding resource situation allowed timely completion. Yet, looking at the instantiation complexity and completion time of process instances in a sliding time window does not provide sufficient information whether bottlenecks occur within the process. This information is obtained by calculating the difference between the number of completed instances and the number of started instances during a control cycle. This value forms the change in backlog and indicates whether the number of currently active process instances has increased or decreased. The number of active VM indicates the amount of computational resources and is determined by the number of VMs actually in service behind a load balancer.

Starting and stopping VMs takes a certain amount of time, simply registering or deregistering a running VM with a load balancer is considerably faster. Also, a VM may be charged in billing time units such as by the hour counting from the start time of the VM. Economic considerations therefore dictate that a VM should be stopped close to the end of the billing time unit.

Scaling-in is intended to reduce the number of VMs in service behind a load balancer by a certain number. This can be achieved by stopping or merely deregistering VMs. Scaling-out is the task of increasing the number of VMs in service behind a load balancer. Registering idle VMs is preferred over starting and registering new VMs. Hence, the first step is to identify the set of running, yet idle and not registered VMs. We then take as many required VMs as possible from this set and re-register them. If necessary, the number of additionally needed VMs is started concurrently.

A third option is that no scaling is needed, i.e., to not change the current distribution of VMs. Hence, the controller performs no action and eventually starts the next control cycle.

\subsection{The Vienna Platform for Elastic Processes}
\label{sub:viepep}
\begin{figure}[htb]
	\centering
	\includegraphics[width=\linewidth]{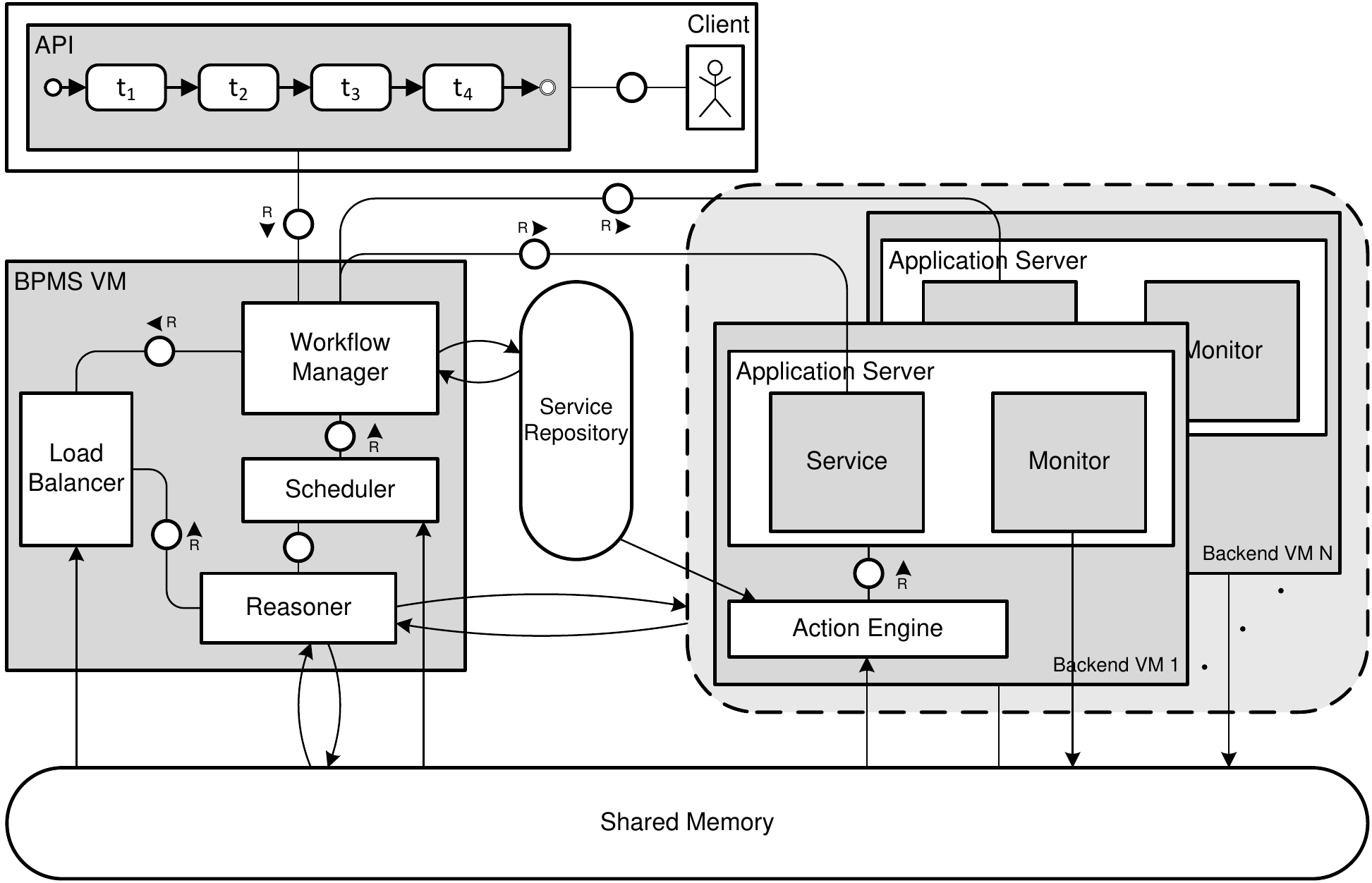}
	\caption{The Vienna Platform for Elastic Processes (ViePEP) -- Overview}
	\label{fig:viepep}
\end{figure}

The \textit{Vienna Platform for Elastic Processes} (ViePEP) is an eBPMS implemented for usage with Openstack\footnote{\url{https://www.openstack.org/}} \cite{hoenisch13,hoenisch13a,schulte13a}. Figure~\ref{fig:viepep} provides an overview of the ViePEP architecture. 

ViePEP takes care of hosting and managing software services in a cloud environment and maps process requests (issued by clients, i.e., the process requesters) onto services which are deployed on cloud-based computational resources in order to execute them. ViePEP considers and ensures SLAs stated by clients while still being as cost-efficient as possible. 

In order to realize this, ViePEP provides the following top level components: The \textit{Client} can issue a process request in terms of a process model and input data. In addition, the Client can define SLAs for the process request. Within these SLAs, deadlines on process level or for a particular task can be defined. After receiving the request, ViePEP schedules and performs the process enactment while ensuring the given SLAs are met. 

To achieve this, the \textit{BPMS VM} offers the eBPMS features of ViePEP, as follows. The \textit{Workflow Manager} is responsible for accepting process requests. These requests are stored in a queue for later or immediate enactment. If an SLA is defined for a particular process, the \textit{Scheduler} is able to compute at what point of time the execution has to be started in order to ensure SLA adherence. This start time is computed by the scheduler by making use of a deadline-based scheduling algorithm \cite{hoenisch13a}. The scheduling plan is handed to the \textit{Reasoner} which computes the required resources. If the reasoner derives that a particular VM is under or over-utilized, rescheduling is triggered \cite{hoenisch13}. Beside accepting process requests, the workflow manager is responsible for performing the actual execution, i.e., it assigns each task to a specific software service running on a particular VM. The \textit{Load Balancer} is able to balance the service requests on the available resources in order to have a relatively even load on each VM. The information for that is read from the \textit{Shared Memory} which stores the actual CPU and RAM load on the available resources. This information is also used by the reasoner, which enables ViePEP to react to an under or overprovisioned system in time and to issue corresponding countermeasures, i.e., \textit{Start/Stop} a VM and \textit{Deploy/Move} a service \cite{schulte13}. 

The \textit{Backend VM} is another crucial part of ViePEP. In a ViePEP-enabled elastic process landscape, several \textit{Backend VMs} are deployed. These host an \textit{Application Server} on which a single service is hosted. Besides this service, a \textit{Monitor} is started, which measures the system's CPU and RAM usage and the deployed service's response time. It should be noted that ViePEP's current monitoring approach is based on CEP. Monitoring data is stored and shared between the BPMS VM in the shared memory instead. In addition to the application server, an \textit{Action Engine} is deployed. It is able to execute the different actions issued by the reasoner. 

\section{Future Research Directions}
\label{sec:future}
In the previous sections, we have discussed the current state of the art regarding the identified infrastructural challenges for elastic processes, namely scheduling, resource allocation, process monitoring, decentralized coordination, and state management. Furthermore, we have discussed the capabilities of two state of the art research eBPMS prototypes. We found that solutions aiming at elastic processes and elastic BPM are scarce and that the field is in general still in its infancy -- current solutions should be regarded as preliminary and limited early attempts. 

Thus, a broad range of research topics needs to be addressed. In this section, we highlight selected research topics for the different infrastructural challenges discussed in Section~\ref{sec:challenges}. Note that this list is by no means exhaustive -- the list of identified research topics is giving our subjective impression of the most urgent and relevant research challenges at the time of writing.

\subsection{Challenges 1 \& 2: Scheduling and Resource Allocation}
\label{sub:scheduling}
We strongly believe that scheduling and resource allocation for elastic processes should always be seen together, since open research questions need to regarded for both challenges. 

\paragraph{Hybrid Clouds and Interclouds} So far, no approach for elastic processes takes into account hybrid clouds, i.e., a combination of computational resources from private and public clouds \cite{Mell2011}, or interclouds, i.e., the usage of multiple clouds from different providers \cite{buyya10}. However, outsourcing of particular tasks from an internal data center to a public cloud is deemed to be a major application area for future cloud computing \cite{vdbossche13}, and usage of multiple clouds in general is a common means to spread risk and circumvent the vendor lock-in issue. Hence, future work on elastic processes should be able to take into account the distinctive features of hybrid clouds or interclouds.
\paragraph{Elasticity Dimensions} All analyzed approaches to scheduling and resource allocation apply \textit{resource} elasticity, while \textit{quality} elasticity is partially regarded in some approaches through optimization of turnaround time of elastic processes. However, other quality aspects like reliability or availability are not regarded at all, and even deadlines for particular process instances are not supported by most scheduling approaches discussed in Section~\ref{sub:processcontrol}. Since timeliness is a critical factor in business processes, quality elasticity should be considered in elastic process scheduling. Furthermore, there are several cloud providers who offer elastic prices, with Amazon's spot instances being the most popular example. Hence, elastic process scheduling should also be able to take into account \textit{cost} elasticity. 	
\paragraph{Data Transfer} Data transfer to and from a VM could lead to additional cost for two reasons: First, many cloud providers charge additional fees for up- and downloading data to and from their cloud infrastructures. Second, based on the provided bandwidth, data transfer may lead to additional VM leasing time. While this is a well-known issue in SWFs \cite{pandey10}, elastic process scheduling and resource allocation do not necessarily take into account cost related to data transfer. 	
\paragraph{Testbeds} Most of the analyzed approaches for scheduling and resource allocation only provide simulations, with the frameworks presented in Section~\ref{sec:realisations} being notable exceptions \cite{hoenisch13,hoenisch13a,euting14}. Naturally, a simulator always abstracts from the real world, which may substantially reduce the significance of the proposed solutions, since there are numerous reasons why the performance of cloud computational resources may vary \cite{Schad10}. Hence, scheduling and resource allocation approaches should be developed and tested in testbeds.
\paragraph{Horizontal vs. Vertical Scaling} Current approaches focus on horizontal scaling. However, further optimization may be achieved by also supporting vertical scaling, since it could lead to lower cost when leasing and releasing computational resources. Furthermore, some cloud providers support vertical scaling at runtime, which can lead to increased speed in realizing scaling decisions. Hence, scheduling and resource allocation approaches which are able to support horizontal and vertical scaling at the same time need to be developed.
\paragraph{Design Methodologies} Designing an eBPMS in the cloud is a complex task which requires the design of processes, services, virtual infrastructure, and scaling algorithms as well as dealing with interdependencies between the different aspects. As of now, there is no synchronized design methodology to model processes, events, and infrastructure. However, this is a prerequisite for realizing elastic process landscapes.
\paragraph{KPIs for Scaling Decisions} In traditional cloud setups, scaling decisions have been taken mostly in a reactive fashion based on local knowledge such as CPU, storage or RAM utilization. For elastic BPM, new KPIs for processes, tasks, and infrastructure have to be developed to analyze the situation and derive meaningful global scaling decisions across ASGs. Examples include changes in process backlogs, process complexity, and average process turnaround time \cite{euting14}.

\subsection{Challenge 3: Process Monitoring and Data Collection}
\label{sub:challenge3}
\paragraph{Common Event Format} Neither for BPM \cite{becker12} nor in the context of cloud computing there is a common, standardized event format which can be used out of the box. In BPM, there is a tendency to make events interoperable using formats such as the Business Process Analytics Format (BPAF) \cite{wfmc08}. In addition, efforts in the IEEE CIS Task Force on Process Mining have led to the upcoming Extensible Event Stream (XES) standard \cite{guentherverbeek14} suitable for process mining \cite{aalst11}, which does not consider the specificities of CEP such as single instance processing, flat rather than nested structures, inclusion of business data, etc.

Cloud computing is still in its interoperability infancy. There is no common format or syntax to analyze the performance and/or states of federated cloud resources at all as cloud providers tend to only provide proprietary access to monitoring information.
\paragraph{Common Event Types} Again, neither for BPM nor in the context of cloud computing there is a standardized set of event types for lifecycle events and business events which can be used to design system interaction. A common set of events would alleviate interoperability issues on the semantic, rather than on the syntactic level.
\paragraph{Common Event Triggers} On the one hand, events can be used to derive higher-level knowledge through CEP, which can then be displayed on dashboards or as alerts. On the other hand, events can be used to initiate an insight-to-action feedback loop. Yet neither for BPM nor for cloud infrastructure there is a standardized set of triggers or a standardized interface to trigger such actions.

\subsection{Challenge 4: Decentralized Coordination for Process Enactment}
\label{sub:challenge4}
\paragraph{Decomposition/Decoupling Processes} Decentralized models of eBPMS would work best when the tasks of elastic processes are cleanly decoupled from each other. This implies that task dependencies have to be carefully analyzed and highly-dependent tasks clustered together. This impacts the scheduling and resource allocation mechanisms employed as well.  
\paragraph{Task Coordination} Coordination mechanisms in distributed systems are well-established. However, there is a need to extend them into richer task coordination mechanisms for decentralized eBPMS. Such mechanisms must be able to support a more detailed vocabulary for communication between the different parts without the cost of increased message complexity.

\subsection{Challenge 5: State Management}
\label{sub:challenge5}
\paragraph{Configuration Management} As discussed in Section~\ref{sub:statemanagement}, configuration management is essential for smooth elasticity. In a dynamic environment, it may be impossible to keep track of all the configuration changes. However, there is a need for reliable failure detectors that can detect harmful configuration changes and repair them before they propagate throughout the system.
\paragraph{Common Cloud State Model} While for BPM there are several state models available which are part of standard specifications, e.g., \cite{wfmc98}, for cloud computing, state models for virtualized infrastructure are neither standardized nor advertised. They may or may not be derivable from documentation, but there is no common cloud state model which can be used to judge the adequacy of a VM for business process enactment.
\paragraph{State Awareness and State Synchronization}
In some of our experiments \cite{janiesch14,euting14}, we observed issues with scale-down decisions of non-process-aware auto-scaling controllers, such as shutting down VMs that are still actively processing requests. One possible interpretation of this behavior is that standard auto-scalers are optimized for Web applications, where user requests typically need to be answered in sub-second time spans. If, in contrast, a computation-intensive service request takes a minute to be answered, running requests will be interrupted by the auto-scaler. 
For instance, the standard AWS policy for scale-in is to shutdown the VM that has been running the longest, not the one that is least busy.
As argued in Section~\ref{sub:allocation}, scale-in decisions should be executed in a way that allows running requests to complete.

Similarly, shutting down VMs hosting eBPMS instances can be a tricky issue. In particular, a eBPMS may be waiting for services to answer requests of hundreds of process instances simultaneously. Thus, even when the CPU of the VM is largely idle, the eBPMS-level load on the machine can be fairly high. A state-unaware scaler may decide to shutdown a VM hosting an eBPMS, due to low CPU utilization.

In general, VM state and application state should be synchronized, so as to implement better scale-in mechanisms.

\section{Conclusion}
\label{sec:conclusion}
Elastic BPM is a promising approach to realize scalable business processes in a cost-efficient and flexible way. Since elastic BPM constitutes a relatively new field of research, the number of respective solutions and research approaches is still small.

Within the scope of this paper, we have analyzed the state of the art in this field, with a focus on the infrastructural challenges of elastic BPM and elastic processes, namely scheduling, resource allocation, process monitoring, decentralized coordination, and state management. Furthermore, two state of the art eBPMS prototypes have been presented, namely the \textit{Fuzzy BPM-aware Auto-Scaler} by KIT and NICTA and the \textit{Vienna Platform for Elastic Processes} by TU Vienna. 

Our analysis has shown that state of the art approaches provide support for some of the single infrastructural challenges, but holistic solutions are not available yet. In addition, there is a large number of research questions where solutions are so far not available at all. Only if these issues will be solved, potential participants in elastic processes will be encouraged to apply such technologies and realize the potential benefits of elastic processes and elastic BPM.

We expect elastic BPM to have a major impact on both the BPM and the cloud computing research communities. Apart from the technological infrastructural challenges we have focused on, future research may take into account more business- and application-driven research questions and analyze the applicability of elastic processes in different application domains.

\section*{Acknowledgements}
This paper is supported by TU Vienna research funds and by the Austrian Science Fund (FWF): P23313-N23. 
This work is supported by the German Academic Exchange Service (DAAD) under the promotional reference ``54392100'' and by the Group of Eight (Go8) Australia-Germany Joint Research Cooperation Scheme.
NICTA is funded by the Australian Government as represented by the Department of Broadband, Communications and the Digital Economy and the Australian Research Council through the ICT Centre of Excellence program.





\section*{References}

All Websites listed in this paper have been last checked on July 17th, 2014.
\bibliographystyle{model1-num-names}
\bibliography{biblio}

\begin{thebibliography}{101}
\expandafter\ifx\csname natexlab\endcsname\relax\def\natexlab#1{#1}\fi
\providecommand{\url}[1]{\texttt{#1}}
\providecommand{\href}[2]{#2}
\providecommand{\path}[1]{#1}
\providecommand{\DOIprefix}{doi:}
\providecommand{\ArXivprefix}{arXiv:}
\providecommand{\URLprefix}{URL: }
\providecommand{\Pubmedprefix}{pmid:}
\providecommand{\doi}[1]{\href{http://dx.doi.org/#1}{\path{#1}}}
\providecommand{\Pubmed}[1]{\href{pmid:#1}{\path{#1}}}
\providecommand{\bibinfo}[2]{#2}
\ifx\xfnm\relax \def\xfnm[#1]{\unskip,\space#1}\fi
\bibitem[{Breu et~al.(2013)Breu, Dustdar, Eder, Huemer, Kappel, K\"{o}pke,
  Langer, Mangler, Mendling, Neumann, Rinderle-Ma, Schulte, Sobernig, and
  Weber}]{breu13}
\bibinfo{author}{R.~Breu}, \bibinfo{author}{S.~Dustdar},
  \bibinfo{author}{J.~Eder}, \bibinfo{author}{C.~Huemer},
  \bibinfo{author}{G.~Kappel}, \bibinfo{author}{J.~K\"{o}pke},
  \bibinfo{author}{P.~Langer}, \bibinfo{author}{J.~Mangler},
  \bibinfo{author}{J.~Mendling}, \bibinfo{author}{G.~Neumann},
  \bibinfo{author}{S.~Rinderle-Ma}, \bibinfo{author}{S.~Schulte},
  \bibinfo{author}{S.~Sobernig}, \bibinfo{author}{B.~Weber},
\newblock \bibinfo{title}{{Towards Living Inter-Organizational Processes}},
\newblock in: \bibinfo{booktitle}{15th IEEE Conference on Business Informatics
  (CBI 2013)}, \bibinfo{publisher}{IEEE}, \bibinfo{year}{2013}, pp.
  \bibinfo{pages}{363--366}.
\bibitem[{Armbrust et~al.(2010)Armbrust, Fox, Griffith, Joseph, Katz,
  Konwinski, Lee, Patterson, Rabkin, Stoica, and Zaharia}]{Armbrust10}
\bibinfo{author}{M.~Armbrust}, \bibinfo{author}{A.~Fox},
  \bibinfo{author}{R.~Griffith}, \bibinfo{author}{A.~D. Joseph},
  \bibinfo{author}{R.~Katz}, \bibinfo{author}{A.~Konwinski},
  \bibinfo{author}{G.~Lee}, \bibinfo{author}{D.~Patterson},
  \bibinfo{author}{A.~Rabkin}, \bibinfo{author}{I.~Stoica},
  \bibinfo{author}{M.~Zaharia},
\newblock \bibinfo{title}{{A View of Cloud Computing}},
\newblock \bibinfo{journal}{Communications of the ACM} \bibinfo{volume}{53}
  (\bibinfo{year}{2010}) \bibinfo{pages}{50--58}.
\bibitem[{Barker and {van Hemert}(2008)}]{bark08}
\bibinfo{author}{A.~Barker}, \bibinfo{author}{J.~{van Hemert}},
\newblock \bibinfo{title}{Scientific workflow: A survey and research
  directions},
\newblock in: \bibinfo{booktitle}{7th International Conference on Parallel
  Processing and Applied Mathematics (PPAM 2008)}, volume
  \bibinfo{volume}{4967} of \textit{\bibinfo{series}{LNCS}},
  \bibinfo{publisher}{Springer}, \bibinfo{year}{2008}, pp.
  \bibinfo{pages}{746--753}.
\bibitem[{V{\"o}ckler et~al.(2011)V{\"o}ckler, Juve, Deelman, Rynge, and
  Berriman}]{voec11}
\bibinfo{author}{J.-S. V{\"o}ckler}, \bibinfo{author}{G.~Juve},
  \bibinfo{author}{E.~Deelman}, \bibinfo{author}{M.~Rynge},
  \bibinfo{author}{G.~B. Berriman},
\newblock \bibinfo{title}{Experiences using cloud computing for a scientific
  workflow application},
\newblock in: \bibinfo{booktitle}{2nd International Workshop on Scientific
  Cloud Computing (ScienceCloud 2011)}, \bibinfo{publisher}{ACM},
  \bibinfo{year}{2011}, pp. \bibinfo{pages}{15--24}.
\bibitem[{Dustdar et~al.(2011)Dustdar, Guo, Satzger, and Truong}]{dustdar11}
\bibinfo{author}{S.~Dustdar}, \bibinfo{author}{Y.~Guo},
  \bibinfo{author}{B.~Satzger}, \bibinfo{author}{H.~L. Truong},
\newblock \bibinfo{title}{{Principles of Elastic Processes}},
\newblock \bibinfo{journal}{IEEE Internet Computing} \bibinfo{volume}{15}
  (\bibinfo{year}{2011}) \bibinfo{pages}{66--71}.
\bibitem[{Weske(2012)}]{Weske2012}
\bibinfo{author}{M.~Weske}, \bibinfo{title}{{Business Process Management:
  Concepts, Languages, Architectures}}, \bibinfo{edition}{2nd} ed.,
  \bibinfo{publisher}{Springer}, \bibinfo{year}{2012}.
\bibitem[{{Object Management Group}(2009)}]{omg11}
\bibinfo{author}{{Object Management Group}}, \bibinfo{title}{{Business Process
  Model and Notation (BPMN) Version 1.2}}, \bibinfo{year}{2009}.
\bibitem[{zur Muehlen and Indulska(2010)}]{zurmuehlen10}
\bibinfo{author}{M.~zur Muehlen}, \bibinfo{author}{M.~Indulska},
\newblock \bibinfo{title}{{Modeling languages for business processes and
  business rules: A representational analysis}},
\newblock \bibinfo{journal}{Information Systems} \bibinfo{volume}{35}
  (\bibinfo{year}{2010}) \bibinfo{pages}{379--390}.
\bibitem[{Davenport(1993)}]{Davenport1993}
\bibinfo{author}{T.~H. Davenport}, \bibinfo{title}{{Process Innovation:
  Reengineering Work Through Information Technology}},
  \bibinfo{publisher}{Harvard Business School Press, Boston, MA},
  \bibinfo{year}{1993}.
\bibitem[{Becker et~al.(2011)Becker, Kugeler, and
  Rosemann}]{BeckerKugelerRosemann2011}
\bibinfo{editor}{J.~Becker}, \bibinfo{editor}{M.~Kugeler},
  \bibinfo{editor}{M.~Rosemann} (Eds.), \bibinfo{title}{{Process Management: A
  Guide for the Design of Business Processes}}, \bibinfo{publisher}{Springer},
  \bibinfo{year}{2011}.
\bibitem[{{van der Aalst} et~al.(2003){van der Aalst}, {ter Hofstede}, and
  Weske}]{VanDerAalst2003}
\bibinfo{author}{W.~M.~P. {van der Aalst}}, \bibinfo{author}{A.~H.~M. {ter
  Hofstede}}, \bibinfo{author}{M.~Weske},
\newblock \bibinfo{title}{{Business process management: A Survey}},
\newblock in: \bibinfo{booktitle}{International Conference on Business Process
  Management (BPM 2003)}, volume \bibinfo{volume}{2678} of
  \textit{\bibinfo{series}{LNCS}}, \bibinfo{publisher}{Springer},
  \bibinfo{year}{2003}, pp. \bibinfo{pages}{1--12}.
\bibitem[{Lud{\"a}scher et~al.(2009)Lud{\"a}scher, Weske, McPhillips, and
  Bowers}]{ludaescher09}
\bibinfo{author}{B.~Lud{\"a}scher}, \bibinfo{author}{M.~Weske},
  \bibinfo{author}{T.~M. McPhillips}, \bibinfo{author}{S.~Bowers},
\newblock \bibinfo{title}{{Scientific Workflows: Business as Usual?}},
\newblock in: \bibinfo{booktitle}{7th Internernational Conference on Business
  Process Management (BPM 2009)}, volume \bibinfo{volume}{5701} of
  \textit{\bibinfo{series}{LNCS}}, \bibinfo{publisher}{Springer},
  \bibinfo{year}{2009}, pp. \bibinfo{pages}{31--47}.
\bibitem[{Mell and Grance(2011)}]{Mell2011}
\bibinfo{author}{P.~Mell}, \bibinfo{author}{T.~Grance}, \bibinfo{title}{{The
  NIST Definition of Cloud Computing}}, \bibinfo{publisher}{Recommendations of
  the National Institute of Standards and Technology}, \bibinfo{year}{2011}.
\bibitem[{C\'{a}ceres et~al.(2010)C\'{a}ceres, Vaquero, Rodero-Merino, {\'{A}.
  Polo}, and Hierro}]{caceres10}
\bibinfo{author}{J.~C\'{a}ceres}, \bibinfo{author}{L.~M. Vaquero},
  \bibinfo{author}{L.~Rodero-Merino}, \bibinfo{author}{{\'{A}. Polo}},
  \bibinfo{author}{J.~J. Hierro},
\newblock \bibinfo{title}{{Service Scalability Over the Cloud}},
\newblock in: \bibinfo{editor}{B.~Furht}, \bibinfo{editor}{A.~Escalante}
  (Eds.), \bibinfo{booktitle}{Handbook of Cloud Computing},
  \bibinfo{publisher}{Springer}, \bibinfo{year}{2010}, pp.
  \bibinfo{pages}{357--377}.
\bibitem[{Copil et~al.(2013)Copil, Moldovan, Truong, and Dustdar}]{copil13}
\bibinfo{author}{G.~Copil}, \bibinfo{author}{D.~Moldovan},
  \bibinfo{author}{H.~L. Truong}, \bibinfo{author}{S.~Dustdar},
\newblock \bibinfo{title}{Multi-level elasticity control of cloud services},
\newblock in: \bibinfo{booktitle}{11th International Conference on
  Service-Oriented Computing (ICSOC 2013)}, volume \bibinfo{volume}{8274} of
  \textit{\bibinfo{series}{LNCS}}, \bibinfo{publisher}{Springer},
  \bibinfo{year}{2013}, pp. \bibinfo{pages}{429--436}.
\bibitem[{Mans et~al.(2010)Mans, Russell, {van der Aalst}, Moleman, and
  Bakker}]{mans10}
\bibinfo{author}{R.~Mans}, \bibinfo{author}{N.~C. Russell},
  \bibinfo{author}{W.~M.~P. {van der Aalst}}, \bibinfo{author}{A.~J. Moleman},
  \bibinfo{author}{P.~J.~M. Bakker},
\newblock \bibinfo{title}{{Schedule-Aware Workflow Management Systems}},
\newblock \bibinfo{journal}{Transactions on Petri Nets and Other Models of
  Concurrency IV} \bibinfo{volume}{6550} (\bibinfo{year}{2010})
  \bibinfo{pages}{121--143}.
\bibitem[{Schulte et~al.(2014)Schulte, Hoenisch, Hochreiner, Dustdar, Klusch,
  and Schuller}]{schulte14}
\bibinfo{author}{S.~Schulte}, \bibinfo{author}{P.~Hoenisch},
  \bibinfo{author}{C.~Hochreiner}, \bibinfo{author}{S.~Dustdar},
  \bibinfo{author}{M.~Klusch}, \bibinfo{author}{D.~Schuller},
\newblock \bibinfo{title}{{Towards Process Support for Cloud Manufacturing
  (accepted for publication)}},
\newblock in: \bibinfo{booktitle}{18th IEEE International Enterprise
  Distributed Object Computing Conference (EDOC 2014)},
  \bibinfo{publisher}{IEEE Computer Society, Washington, DC, USA},
  \bibinfo{year}{2014}, pp. \bibinfo{pages}{NN--NN}.
\bibitem[{Rohjans et~al.(2012)Rohjans, D{\"a}nekas, and Uslar}]{rohjans12}
\bibinfo{author}{S.~Rohjans}, \bibinfo{author}{C.~D{\"a}nekas},
  \bibinfo{author}{M.~Uslar},
\newblock \bibinfo{title}{{Requirements for Smart Grid ICT Architectures}},
\newblock in: \bibinfo{booktitle}{Third IEEE PES Innovative Smart Grid
  Technologies (ISGT) Europe Conf.}, \bibinfo{publisher}{IEEE},
  \bibinfo{year}{2012}, pp. \bibinfo{pages}{1--8}.
\bibitem[{Lampe et~al.(2013)Lampe, Wenge, M\"{u}ller, and
  Schaarschmidt}]{lampe13}
\bibinfo{author}{U.~Lampe}, \bibinfo{author}{O.~Wenge},
  \bibinfo{author}{A.~M\"{u}ller}, \bibinfo{author}{R.~Schaarschmidt},
\newblock \bibinfo{title}{{On the Relevance of Security Risks for Cloud
  Adoption in the Financial Industry}},
\newblock in: \bibinfo{booktitle}{19th Americas Conference on Information
  Systems (AMCIS 2013)}, \bibinfo{publisher}{AIS}, \bibinfo{year}{2013}.
\bibitem[{Gill et~al.(2011)Gill, Bunker, and Seltsikas}]{gill11}
\bibinfo{author}{A.~Q. Gill}, \bibinfo{author}{D.~Bunker},
  \bibinfo{author}{P.~Seltsikas},
\newblock \bibinfo{title}{{An Empirical Analysis of Cloud, Mobile, Social and
  Green Computing -- Financial Services IT Strategy and Enterprise
  Architecture}},
\newblock in: \bibinfo{booktitle}{IEEE Ninth International Conference on
  Dependable, Autonomic and Secure Computing (DASC 2011)},
  \bibinfo{publisher}{IEEE}, \bibinfo{year}{2011}, pp.
  \bibinfo{pages}{697--704}.
\bibitem[{Rabhi et~al.(2012)Rabhi, Yao, and Guabtni}]{rabhi12}
\bibinfo{author}{F.~A. Rabhi}, \bibinfo{author}{L.~Yao},
  \bibinfo{author}{A.~Guabtni},
\newblock \bibinfo{title}{{ADAGE: a framework for supporting user-driven ad-hoc
  data analysis processes}},
\newblock \bibinfo{journal}{Computing} \bibinfo{volume}{94}
  (\bibinfo{year}{2012}) \bibinfo{pages}{489--519}.
\bibitem[{Shi et~al.(2010)Shi, Xia, and Zhan}]{shi10}
\bibinfo{author}{A.~Shi}, \bibinfo{author}{Y.~Xia}, \bibinfo{author}{H.~Zhan},
\newblock \bibinfo{title}{{Applying Cloud Computing in Financial Service
  Industry}},
\newblock in: \bibinfo{booktitle}{International Conference on Intelligent
  Control and Information Processing}, \bibinfo{publisher}{IEEE},
  \bibinfo{year}{2010}, pp. \bibinfo{pages}{579--583}.
\bibitem[{Janiesch et~al.(2014)Janiesch, Weber, Kuhlenkamp, and
  Menzel}]{janiesch14}
\bibinfo{author}{C.~Janiesch}, \bibinfo{author}{I.~Weber},
  \bibinfo{author}{J.~Kuhlenkamp}, \bibinfo{author}{M.~Menzel},
\newblock \bibinfo{title}{{Optimizing the Performance of Automated Business
  Processes Executed on Virtualized Infrastructure}},
\newblock in: \bibinfo{booktitle}{47th Hawaii International Conference on
  System Sciences (HICSS 2014)}, \bibinfo{publisher}{IEEE},
  \bibinfo{year}{2014}, pp. \bibinfo{pages}{3818--3826}.
\bibitem[{Eder et~al.(1999)Eder, Panagos, and Rabinovich}]{eder99}
\bibinfo{author}{J.~Eder}, \bibinfo{author}{E.~Panagos},
  \bibinfo{author}{M.~Rabinovich},
\newblock \bibinfo{title}{{Time Constraints in Workflow Systems}},
\newblock in: \bibinfo{booktitle}{11th International Conference on Advanced
  Information Systems Engineering (CAiSE'99)}, volume \bibinfo{volume}{1626} of
  \textit{\bibinfo{series}{LNCS}}, \bibinfo{publisher}{Springer},
  \bibinfo{year}{1999}, pp. \bibinfo{pages}{286--300}.
\bibitem[{Avanes and Freytag(2008)}]{avanes08}
\bibinfo{author}{A.~Avanes}, \bibinfo{author}{J.~C. Freytag},
\newblock \bibinfo{title}{{Adaptive Workflow Scheduling Under Resource
  Allocation Constraints and Network Dynamics}},
\newblock \bibinfo{journal}{Proceedings of the VLDB Endowment}
  \bibinfo{volume}{1} (\bibinfo{year}{2008}) \bibinfo{pages}{1631--1637}.
\bibitem[{Huang et~al.(2011)Huang, {van der Aalst}, Lu, and Duan}]{huang11}
\bibinfo{author}{Z.~Huang}, \bibinfo{author}{W.~M.~P. {van der Aalst}},
  \bibinfo{author}{X.~Lu}, \bibinfo{author}{H.~Duan},
\newblock \bibinfo{title}{{Reinforcement learning based resource allocation in
  business process management}},
\newblock \bibinfo{journal}{Data \& Knowledge Engineering} \bibinfo{volume}{70}
  (\bibinfo{year}{2011}) \bibinfo{pages}{127--145}.
\bibitem[{Yu et~al.(2008)Yu, Buyya, and Ramamohanarao}]{yu08a}
\bibinfo{author}{J.~Yu}, \bibinfo{author}{R.~Buyya},
  \bibinfo{author}{K.~Ramamohanarao},
\newblock \bibinfo{title}{{Workflow Scheduling Algorithms for Grid Computing}},
\newblock in: \bibinfo{editor}{F.~Xhafa}, \bibinfo{editor}{A.~Abraham} (Eds.),
  \bibinfo{booktitle}{{Metaheuristics for Scheduling in Distributed Computing
  Environments}}, volume \bibinfo{volume}{146} of
  \textit{\bibinfo{series}{Studies in Computational Intelligence}},
  \bibinfo{publisher}{Springer}, \bibinfo{year}{2008}, pp.
  \bibinfo{pages}{173--214}.
\bibitem[{Baker and Trietsch(2009)}]{baker09}
\bibinfo{author}{K.~R. Baker}, \bibinfo{author}{D.~Trietsch},
  \bibinfo{title}{{Principles of Sequencing and Scheduling}},
  \bibinfo{publisher}{John Wiley \& Sons, Hoboken, NJ}, \bibinfo{year}{2009}.
\bibitem[{Strunk(2010)}]{strunk10}
\bibinfo{author}{A.~Strunk},
\newblock \bibinfo{title}{{Q}o{S}-{A}ware {S}ervice {C}omposition: {A}
  {S}urvey},
\newblock in: \bibinfo{booktitle}{8th IEEE European Conference on Web Services
  (ECOWS 2010)}, \bibinfo{publisher}{IEEE}, \bibinfo{year}{2010}, pp.
  \bibinfo{pages}{67--74}.
\bibitem[{Garey and Johnson(1979)}]{garey79}
\bibinfo{author}{M.~R. Garey}, \bibinfo{author}{D.~S. Johnson},
  \bibinfo{title}{{Computers and Intractability: A Guide to the Theory of
  NP-Completeness}}, \bibinfo{publisher}{W.H. Freeman \& Co., New York, NY,
  USA}, \bibinfo{year}{1979}.
\bibitem[{Schad et~al.(2010)Schad, Dittrich, and Quian{\'e}-Ruiz}]{Schad10}
\bibinfo{author}{J.~Schad}, \bibinfo{author}{J.~Dittrich},
  \bibinfo{author}{J.-A. Quian{\'e}-Ruiz},
\newblock \bibinfo{title}{Runtime measurements in the cloud: Observing,
  analyzing, and reducing variance},
\newblock \bibinfo{journal}{Proceedings of the VLDB Endowment}
  \bibinfo{volume}{3} (\bibinfo{year}{2010}) \bibinfo{pages}{460--471}.
\bibitem[{Baggio et~al.(2004)Baggio, Wainer, and Ellis}]{baggio04}
\bibinfo{author}{G.~Baggio}, \bibinfo{author}{J.~Wainer},
  \bibinfo{author}{C.~A. Ellis},
\newblock \bibinfo{title}{Applying scheduling techniques to minimize the number
  of late jobs in workflow systems},
\newblock in: \bibinfo{booktitle}{19th Symposium on Applied Computing (SAC
  2004)}, \bibinfo{publisher}{ACM}, \bibinfo{year}{2004}, pp.
  \bibinfo{pages}{1396--1403}.
\bibitem[{Pinedo(2012)}]{pinedo12}
\bibinfo{author}{M.~L. Pinedo}, \bibinfo{title}{Scheduling -- Algorithms, and
  Systems}, \bibinfo{edition}{4th} ed., \bibinfo{publisher}{Springer, New
  York}, \bibinfo{year}{2012}.
\bibitem[{Combi and Pozzi(2006)}]{combi06}
\bibinfo{author}{C.~Combi}, \bibinfo{author}{G.~Pozzi},
\newblock \bibinfo{title}{{Task Scheduling for a Temporal Workflow Management
  System}},
\newblock in: \bibinfo{booktitle}{13th International Symposium on Temporal
  Representation and Reasoning (TIME'06)}, \bibinfo{publisher}{IEEE},
  \bibinfo{year}{2006}, pp. \bibinfo{pages}{61--68}.
\bibitem[{Lanz et~al.(2012)Lanz, Weber, and Reichert}]{lanz12}
\bibinfo{author}{A.~Lanz}, \bibinfo{author}{B.~Weber},
  \bibinfo{author}{M.~Reichert},
\newblock \bibinfo{title}{Time patterns for process-aware information systems},
\newblock \bibinfo{journal}{Requirements Engineering}  (\bibinfo{year}{2012})
  \bibinfo{pages}{1--29}.
\bibitem[{Senkul and Toroslu(2005)}]{senkul05}
\bibinfo{author}{P.~Senkul}, \bibinfo{author}{I.~H. Toroslu},
\newblock \bibinfo{title}{{An Architecture for Workflow Scheduling Under
  Resource Allocation Constraints}},
\newblock \bibinfo{journal}{Information Systems} \bibinfo{volume}{30}
  (\bibinfo{year}{2005}) \bibinfo{pages}{399--422}.
\bibitem[{Fr\^{\i}ncu et~al.(2013)Fr\^{\i}ncu, Genaud, and Gossa}]{frincu13}
\bibinfo{author}{M.~E. Fr\^{\i}ncu}, \bibinfo{author}{S.~Genaud},
  \bibinfo{author}{J.~Gossa},
\newblock \bibinfo{title}{{Comparing Provisioning and Scheduling Strategies for
  Workflows on Clouds}},
\newblock in: \bibinfo{booktitle}{2013 IEEE 27th International Symposium on
  Parallel and Distributed Processing (IPDPS 2013) Workshops and PhD Forum},
  \bibinfo{publisher}{IEEE}, \bibinfo{year}{2013}, pp.
  \bibinfo{pages}{2101--2110}.
\bibitem[{Pandey et~al.(2010)Pandey, Wu, Guru, and Buyya}]{pandey10}
\bibinfo{author}{S.~Pandey}, \bibinfo{author}{L.~Wu}, \bibinfo{author}{S.~M.
  Guru}, \bibinfo{author}{R.~Buyya},
\newblock \bibinfo{title}{{A Particle Swarm Optimization-Based Heuristic for
  Scheduling Workflow Applications in Cloud Computing Environments}},
\newblock in: \bibinfo{booktitle}{24th IEEE International Conference on
  Advanced Information Networking and Applications (AINA 2010)},
  \bibinfo{publisher}{IEEE}, \bibinfo{year}{2010}, pp.
  \bibinfo{pages}{400--407}.
\bibitem[{Szabo and Kroeger(2012)}]{szabo12}
\bibinfo{author}{C.~Szabo}, \bibinfo{author}{T.~Kroeger},
\newblock \bibinfo{title}{{Evolving Multi-objective Strategies for Task
  Allocation of Scientific Workflows on Public Clouds}},
\newblock in: \bibinfo{booktitle}{IEEE Congress on Evolutionary Computation
  (CEC 2012)}, \bibinfo{publisher}{IEEE}, \bibinfo{year}{2012}, pp.
  \bibinfo{pages}{1--8}.
\bibitem[{Byun et~al.(2011)Byun, Kee, Kim, and Maeng}]{byun11}
\bibinfo{author}{E.-K. Byun}, \bibinfo{author}{Y.-S. Kee},
  \bibinfo{author}{J.-S. Kim}, \bibinfo{author}{S.~Maeng},
\newblock \bibinfo{title}{Cost optimized provisioning of elastic resources for
  application workflows},
\newblock \bibinfo{journal}{Future Generation Computer Systems}
  \bibinfo{volume}{27} (\bibinfo{year}{2011}) \bibinfo{pages}{1011--1026}.
\bibitem[{Abrishami et~al.(2013)Abrishami, Naghibzadeh, and
  Epema}]{abrishami13}
\bibinfo{author}{S.~Abrishami}, \bibinfo{author}{M.~Naghibzadeh},
  \bibinfo{author}{D.~H.~J. Epema},
\newblock \bibinfo{title}{{Deadline-constrained workflow scheduling algorithms
  for Infrastructure as a Service Clouds}},
\newblock \bibinfo{journal}{Future Generation Computer Systems}
  \bibinfo{volume}{29} (\bibinfo{year}{2013}) \bibinfo{pages}{158--169}.
\bibitem[{{van der Aalst} et~al.(2003){van der Aalst}, {ter Hofstede},
  Kiepuszewski, and Barros}]{vdAalst03a}
\bibinfo{author}{W.~M.~P. {van der Aalst}}, \bibinfo{author}{A.~H.~M. {ter
  Hofstede}}, \bibinfo{author}{B.~Kiepuszewski}, \bibinfo{author}{A.~P.
  Barros},
\newblock \bibinfo{title}{{Workflow Patterns}},
\newblock \bibinfo{journal}{Distributed and Parallel Databases}
  \bibinfo{volume}{14} (\bibinfo{year}{2003}) \bibinfo{pages}{5--51}.
\bibitem[{Xu et~al.(2009)Xu, Cui, Wang, and Bi}]{xu09}
\bibinfo{author}{M.~Xu}, \bibinfo{author}{L.-Z. Cui},
  \bibinfo{author}{H.~Wang}, \bibinfo{author}{Y.~Bi},
\newblock \bibinfo{title}{{A Multiple QoS Constrained Scheduling Strategy of
  Multiple Workflows for Cloud Computing}},
\newblock in: \bibinfo{booktitle}{IEEE International Symposium on Parallel and
  Distributed Processing with Applications (ISPA 2009)},
  \bibinfo{publisher}{IEEE}, \bibinfo{year}{2009}, pp.
  \bibinfo{pages}{629--634}.
\bibitem[{Juhnke et~al.(2011)Juhnke, D\"{o}rnemann, Bock, and
  Freisleben}]{juhnke11}
\bibinfo{author}{E.~Juhnke}, \bibinfo{author}{T.~D\"{o}rnemann},
  \bibinfo{author}{D.~Bock}, \bibinfo{author}{B.~Freisleben},
\newblock \bibinfo{title}{{Multi-objective Scheduling of BPEL Workflows in
  Geographically Distributed Clouds}},
\newblock in: \bibinfo{booktitle}{4th International Conference on Cloud
  Computing (CLOUD 2011)}, \bibinfo{publisher}{IEEE}, \bibinfo{year}{2011}, pp.
  \bibinfo{pages}{412--419}.
\bibitem[{Bessai et~al.(2013)Bessai, Youcef, Oulamara, Godart, and
  Nurcan}]{bessai13a}
\bibinfo{author}{K.~Bessai}, \bibinfo{author}{S.~Youcef},
  \bibinfo{author}{A.~Oulamara}, \bibinfo{author}{C.~Godart},
  \bibinfo{author}{S.~Nurcan},
\newblock \bibinfo{title}{{Business Process scheduling strategies in Cloud
  environments with fairness metrics}},
\newblock in: \bibinfo{booktitle}{IEEE 10th International Conference on
  Services Computing (SCC 2013)}, \bibinfo{publisher}{IEEE},
  \bibinfo{year}{2013}, pp. \bibinfo{pages}{519--526}.
\bibitem[{Hoenisch et~al.(2013{\natexlab{a}})Hoenisch, Schulte, Dustdar, and
  Venugopal}]{hoenisch13}
\bibinfo{author}{P.~Hoenisch}, \bibinfo{author}{S.~Schulte},
  \bibinfo{author}{S.~Dustdar}, \bibinfo{author}{S.~Venugopal},
\newblock \bibinfo{title}{{Self-Adaptive Resource Allocation for Elastic
  Process Execution}},
\newblock in: \bibinfo{booktitle}{6th International Conference on Cloud
  Computing (CLOUD 2013)}, \bibinfo{publisher}{IEEE},
  \bibinfo{year}{2013}{\natexlab{a}}, pp. \bibinfo{pages}{220--227}.
\bibitem[{Hoenisch et~al.(2013{\natexlab{b}})Hoenisch, Schulte, and
  Dustdar}]{hoenisch13a}
\bibinfo{author}{P.~Hoenisch}, \bibinfo{author}{S.~Schulte},
  \bibinfo{author}{S.~Dustdar},
\newblock \bibinfo{title}{{Workflow Scheduling and Resource Allocation for
  Cloud-based Execution of Elastic Processes}},
\newblock in: \bibinfo{booktitle}{6th IEEE International Conference on Service
  Oriented Computing and Applications (SOCA 2013)}, \bibinfo{publisher}{IEEE},
  \bibinfo{year}{2013}{\natexlab{b}}, pp. \bibinfo{pages}{1--8}.
\bibitem[{Schulte et~al.(2013)Schulte, Schuller, Hoenisch, Lampe, Dustdar, and
  Steinmetz}]{schulte13a}
\bibinfo{author}{S.~Schulte}, \bibinfo{author}{D.~Schuller},
  \bibinfo{author}{P.~Hoenisch}, \bibinfo{author}{U.~Lampe},
  \bibinfo{author}{S.~Dustdar}, \bibinfo{author}{R.~Steinmetz},
\newblock \bibinfo{title}{{Cost-Driven Optimization of Cloud Resource
  Allocation for Elastic Processes}},
\newblock \bibinfo{journal}{International Journal of Cloud Computing}
  \bibinfo{volume}{1} (\bibinfo{year}{2013}) \bibinfo{pages}{1--14}.
\bibitem[{Leitner et~al.(2012)Leitner, Hummer, Satzger, Inzinger, and
  Dustdar}]{leitner12}
\bibinfo{author}{P.~Leitner}, \bibinfo{author}{W.~Hummer},
  \bibinfo{author}{B.~Satzger}, \bibinfo{author}{C.~Inzinger},
  \bibinfo{author}{S.~Dustdar},
\newblock \bibinfo{title}{{Cost-Efficient and Application SLA-Aware Client Side
  Request Scheduling in an Infrastructure-as-a-Service Cloud}},
\newblock in: \bibinfo{booktitle}{5th International Conference on Cloud
  Computing (CLOUD 2012)}, \bibinfo{publisher}{IEEE}, \bibinfo{year}{2012}, pp.
  \bibinfo{pages}{213--220}.
\bibitem[{Wu et~al.(2011)Wu, Garg, and Buyya}]{wu11}
\bibinfo{author}{L.~Wu}, \bibinfo{author}{S.~K. Garg},
  \bibinfo{author}{R.~Buyya},
\newblock \bibinfo{title}{{SLA-Based Resource Allocation for Software as a
  Service Provider (SaaS) in Cloud Computing Environments}},
\newblock in: \bibinfo{booktitle}{11th IEEE/ACM International Symposium on
  Cluster, Cloud and Grid Computing (CCGrid 2011)}, \bibinfo{publisher}{IEEE},
  \bibinfo{year}{2011}, pp. \bibinfo{pages}{195--204}.
\bibitem[{{Van den Bossche} et~al.(2013){Van den Bossche}, Vanmechelen, and
  Broeckhove}]{vdbossche13}
\bibinfo{author}{R.~{Van den Bossche}}, \bibinfo{author}{K.~Vanmechelen},
  \bibinfo{author}{J.~Broeckhove},
\newblock \bibinfo{title}{{Online cost-efficient scheduling of
  deadline-constrained workloads on hybrid clouds}},
\newblock \bibinfo{journal}{Future Generation Computing Systems}
  \bibinfo{volume}{29} (\bibinfo{year}{2013}) \bibinfo{pages}{973--985}.
\bibitem[{Dietterich(2002)}]{Dietterich02}
\bibinfo{author}{T.~G. Dietterich},
\newblock \bibinfo{title}{Machine learning for sequential data: A review},
\newblock in: \bibinfo{booktitle}{Joint IAPR International Workshop on
  Structural, Syntactic, and Statistical Pattern Recognition (SSPR 2002 and SPR
  2002)}, \bibinfo{year}{2002}, pp. \bibinfo{pages}{15--30}.
\bibitem[{Bermolen and Rossi(2009)}]{Bermolen09}
\bibinfo{author}{P.~Bermolen}, \bibinfo{author}{D.~Rossi},
\newblock \bibinfo{title}{Support vector regression for link load prediction},
\newblock \bibinfo{journal}{Computer Networks} \bibinfo{volume}{53}
  (\bibinfo{year}{2009}) \bibinfo{pages}{191--201}.
\bibitem[{Sladescu et~al.(2012)Sladescu, Fekete, Lee, and Liu}]{Sladescu12}
\bibinfo{author}{M.~Sladescu}, \bibinfo{author}{A.~Fekete},
  \bibinfo{author}{K.~Lee}, \bibinfo{author}{A.~Liu},
\newblock \bibinfo{title}{Event aware workload prediction: A study using
  auction events},
\newblock in: \bibinfo{booktitle}{13th International Conference on Web
  Information Systems Engineering (WISE'12)}, volume \bibinfo{volume}{7651} of
  \textit{\bibinfo{series}{LNCS}}, \bibinfo{publisher}{Springer},
  \bibinfo{year}{2012}, pp. \bibinfo{pages}{368--381}.
\bibitem[{Marzolla and Mirandola(2007)}]{Marzolla07}
\bibinfo{author}{M.~Marzolla}, \bibinfo{author}{R.~Mirandola},
\newblock \bibinfo{title}{Performance prediction of web service workflows},
\newblock in: \bibinfo{booktitle}{Third International Conference on Quality of
  Software Architectures (QoSA 2007)}, volume \bibinfo{volume}{4880} of
  \textit{\bibinfo{series}{LNCS}}, \bibinfo{publisher}{Springer},
  \bibinfo{year}{2007}, pp. \bibinfo{pages}{127--144}.
\bibitem[{Reijers and {van der Aalst}(2005)}]{Reijers05}
\bibinfo{author}{H.~A. Reijers}, \bibinfo{author}{W.~M.~P. {van der Aalst}},
\newblock \bibinfo{title}{The effectiveness of workflow management systems:
  Predictions and lessons learned},
\newblock \bibinfo{journal}{International Journal of Information Management}
  \bibinfo{volume}{25} (\bibinfo{year}{2005}) \bibinfo{pages}{458 -- 472}.
\bibitem[{Heinis et~al.(2005)Heinis, Pautasso, and Alonso}]{Heinis05}
\bibinfo{author}{T.~Heinis}, \bibinfo{author}{C.~Pautasso},
  \bibinfo{author}{G.~Alonso},
\newblock \bibinfo{title}{Design and evaluation of an autonomic workflow
  engine},
\newblock in: \bibinfo{booktitle}{Second International Conference on Autonomic
  Computing (ICAC 2005)}, \bibinfo{year}{2005}, pp. \bibinfo{pages}{27--38}.
\bibitem[{Lampe et~al.(2011)Lampe, Mayer, Hiemer, Schuller, and
  Steinmetz}]{lampe11}
\bibinfo{author}{U.~Lampe}, \bibinfo{author}{T.~Mayer},
  \bibinfo{author}{J.~Hiemer}, \bibinfo{author}{D.~Schuller},
  \bibinfo{author}{R.~Steinmetz},
\newblock \bibinfo{title}{{Enabling Cost-Efficient Software Service
  Distribution in Infrastructure Clouds at Run Time}},
\newblock in: \bibinfo{booktitle}{4th IEEE International Conference on
  Service-Oriented Computing and Applications (SOCA 2011)},
  \bibinfo{publisher}{IEEE}, \bibinfo{year}{2011}, pp. \bibinfo{pages}{1--8}.
\bibitem[{Buyya et~al.(2010)Buyya, Ranjan, and Calheiros}]{buyya10}
\bibinfo{author}{R.~Buyya}, \bibinfo{author}{R.~Ranjan}, \bibinfo{author}{R.~N.
  Calheiros},
\newblock \bibinfo{title}{{InterCloud: Utility-Oriented Federation of Cloud
  Computing Environments for Scaling of Application Services}},
\newblock in: \bibinfo{booktitle}{10th International Conference on Algorithms
  and Architectures for Parallel Processing (ICA3PP 2010)}, volume
  \bibinfo{volume}{6081} of \textit{\bibinfo{series}{LNCS}},
  \bibinfo{publisher}{Springer}, \bibinfo{year}{2010}, pp.
  \bibinfo{pages}{13--31}.
\bibitem[{Cardellini et~al.(2011)Cardellini, Casalicchio, {Lo Presti}, and
  Silvestri}]{Cardellini2011}
\bibinfo{author}{V.~Cardellini}, \bibinfo{author}{E.~Casalicchio},
  \bibinfo{author}{F.~{Lo Presti}}, \bibinfo{author}{L.~Silvestri},
\newblock \bibinfo{title}{{SLA-aware Resource Management for Application
  Service Providers in the Cloud}},
\newblock in: \bibinfo{booktitle}{First International Symposium on Network
  Cloud Computing and Applications (NCCA '11)}, \bibinfo{publisher}{IEEE},
  \bibinfo{year}{2011}, pp. \bibinfo{pages}{20--27}.
\bibitem[{Emeakaroha et~al.(2013)Emeakaroha, Maurer, Stern, Labaj, Brandic, and
  Kreil}]{emeakaroha_managing_2013}
\bibinfo{author}{V.~C. Emeakaroha}, \bibinfo{author}{M.~Maurer},
  \bibinfo{author}{P.~Stern}, \bibinfo{author}{P.~P. Labaj},
  \bibinfo{author}{I.~Brandic}, \bibinfo{author}{D.~P. Kreil},
\newblock \bibinfo{title}{Managing and optimizing bioinformatics workflows for
  data analysis in clouds},
\newblock \bibinfo{journal}{Journal of Grid Computing} \bibinfo{volume}{11}
  (\bibinfo{year}{2013}) \bibinfo{pages}{407--428}.
\bibitem[{Li and Venugopal(2011)}]{li11}
\bibinfo{author}{H.~Li}, \bibinfo{author}{S.~Venugopal},
\newblock \bibinfo{title}{{Using Reinforcement Learning for Controlling an
  Elastic Web Application Hosting Platform}},
\newblock in: \bibinfo{booktitle}{8th International Conference on Autonomic
  Computing (ICAC 2011)}, \bibinfo{publisher}{ACM}, \bibinfo{year}{2011}, pp.
  \bibinfo{pages}{205--208}.
\bibitem[{Euting et~al.(2014)Euting, Janiesch, Fischer, Tai, and
  Weber}]{euting14}
\bibinfo{author}{S.~Euting}, \bibinfo{author}{C.~Janiesch},
  \bibinfo{author}{R.~Fischer}, \bibinfo{author}{S.~Tai},
  \bibinfo{author}{I.~Weber},
\newblock \bibinfo{title}{{Scalable Business Process Execution in the Cloud}},
\newblock in: \bibinfo{booktitle}{2014 IEEE International Conference on Cloud
  Engineering (IC2E 2014)}, \bibinfo{publisher}{IEEE}, \bibinfo{year}{2014},
  pp. \bibinfo{pages}{175--184}.
\bibitem[{Wei and Blake(2013)}]{wei13}
\bibinfo{author}{Y.~Wei}, \bibinfo{author}{M.~B. Blake},
\newblock \bibinfo{title}{{Decentralized Resource Coordination across Service
  Workflows in a Cloud Environment}},
\newblock in: \bibinfo{booktitle}{22nd IEEE International Conference on
  Collaboration Technologies and Infrastructures (WETICE 2013)},
  \bibinfo{publisher}{IEEE}, \bibinfo{year}{2013}, pp. \bibinfo{pages}{15--20}.
\bibitem[{Dejun et~al.(2010)Dejun, Pierre, and Chi}]{dejun10}
\bibinfo{author}{J.~Dejun}, \bibinfo{author}{G.~Pierre}, \bibinfo{author}{C.-H.
  Chi},
\newblock \bibinfo{title}{{Autonomous Resource Provisioning for Multi-service
  Web Applications}},
\newblock in: \bibinfo{booktitle}{19th International Conference on World Wide
  Web (WWW '10)}, \bibinfo{publisher}{ACM}, \bibinfo{year}{2010}, pp.
  \bibinfo{pages}{471--480}.
\bibitem[{Luckham(2002)}]{Luckham2002}
\bibinfo{author}{D.~Luckham}, \bibinfo{title}{{The Power of Events: An
  Introduction to Complex Event Processing in Distributed Enterprise Systems}},
  \bibinfo{publisher}{Addison-Wesley Professional, Boston, MA},
  \bibinfo{year}{2002}.
\bibitem[{Etzion and Niblett(2010)}]{Etzion2010}
\bibinfo{author}{O.~Etzion}, \bibinfo{author}{P.~Niblett},
  \bibinfo{title}{{Event Processing in Action}}, \bibinfo{publisher}{Manning
  Publications, Cincinnati, OH}, \bibinfo{year}{2010}.
\bibitem[{Janiesch et~al.(2012)Janiesch, Matzner, and
  M{\"u}ller}]{Janiesch2012}
\bibinfo{author}{C.~Janiesch}, \bibinfo{author}{M.~Matzner},
  \bibinfo{author}{O.~M{\"u}ller},
\newblock \bibinfo{title}{Beyond process monitoring: a proof-of-concept of
  event-driven business activity management},
\newblock \bibinfo{journal}{Business Process Management J.}
  \bibinfo{volume}{18} (\bibinfo{year}{2012}) \bibinfo{pages}{625--643}.
\bibitem[{Becker et~al.(2012)Becker, Matzner, M{\"u}ller, and
  Walter}]{becker12}
\bibinfo{author}{J.~Becker}, \bibinfo{author}{M.~Matzner},
  \bibinfo{author}{O.~M{\"u}ller}, \bibinfo{author}{M.~Walter},
\newblock \bibinfo{title}{{A Review of Event Formats as Enablers of
  Event-Driven BPM}},
\newblock in: \bibinfo{booktitle}{Business Process Management Workshops (BPM
  2011)}, volume~\bibinfo{volume}{99} of \textit{\bibinfo{series}{LNBIP}},
  \bibinfo{publisher}{Springer}, \bibinfo{year}{2012}, pp.
  \bibinfo{pages}{433--445}.
\bibitem[{Krumeich et~al.(2014)Krumeich, Weis, Werth, and Loos}]{krumeich14}
\bibinfo{author}{J.~Krumeich}, \bibinfo{author}{B.~Weis},
  \bibinfo{author}{D.~Werth}, \bibinfo{author}{P.~Loos},
\newblock \bibinfo{title}{{Event-Driven Business Process Management: where are
  we now?: A comprehensive synthesis and analysis of literature}},
\newblock \bibinfo{journal}{Business Process Management Journal}
  \bibinfo{volume}{20} (\bibinfo{year}{2014}) \bibinfo{pages}{615--633}.
\bibitem[{Alonso et~al.(1995)Alonso, Mohan, G\"{u}nth\"{o}r, Agrawal, {El
  Abbadi}, and Kamath}]{alonso_exotica/fmqm:_1995}
\bibinfo{author}{G.~Alonso}, \bibinfo{author}{C.~Mohan},
  \bibinfo{author}{R.~G\"{u}nth\"{o}r}, \bibinfo{author}{D.~Agrawal},
  \bibinfo{author}{A.~{El Abbadi}}, \bibinfo{author}{M.~Kamath},
\newblock \bibinfo{title}{{Exotica/FMQM:} a persistent message-based
  architecture for distributed workflow management},
\newblock in: \bibinfo{booktitle}{Information Systems Development for
  Decentralized Organizations}, \bibinfo{publisher}{Springer},
  \bibinfo{year}{1995}, pp. \bibinfo{pages}{1--18}.
\bibitem[{Taylor et~al.(2007)Taylor, Shields, Wang, and
  Harrison}]{taylor_triana_2007}
\bibinfo{author}{I.~Taylor}, \bibinfo{author}{M.~Shields},
  \bibinfo{author}{I.~Wang}, \bibinfo{author}{A.~Harrison},
\newblock \bibinfo{title}{The triana workflow environment: Architecture and
  applications},
\newblock in: \bibinfo{editor}{I.~J. Taylor}, \bibinfo{editor}{E.~Deelman},
  \bibinfo{editor}{D.~B. Gannon}, \bibinfo{editor}{M.~Shields} (Eds.),
  \bibinfo{booktitle}{Workflows for e-Science}, \bibinfo{publisher}{Springer
  London}, \bibinfo{year}{2007}, pp. \bibinfo{pages}{320--339}.
\bibitem[{Rahman et~al.(2010)Rahman, Ranjan, and
  Buyya}]{rahman_cooperative_2010}
\bibinfo{author}{M.~Rahman}, \bibinfo{author}{R.~Ranjan},
  \bibinfo{author}{R.~Buyya},
\newblock \bibinfo{title}{Cooperative and decentralized workflow scheduling in
  global grids},
\newblock \bibinfo{journal}{Future Generation Computer Systems}
  \bibinfo{volume}{26} (\bibinfo{year}{2010}) \bibinfo{pages}{753--768}.
\bibitem[{Legrand et~al.(2011)Legrand, Newman, Voicu, Dobre, and
  Grigoras}]{legrand_workflow_2011}
\bibinfo{author}{I.~Legrand}, \bibinfo{author}{H.~Newman},
  \bibinfo{author}{R.~Voicu}, \bibinfo{author}{C.~Dobre},
  \bibinfo{author}{C.~Grigoras},
\newblock \bibinfo{title}{Workflow management in large distributed systems},
\newblock \bibinfo{journal}{Journal of Physics: Conference Series}
  \bibinfo{volume}{331} (\bibinfo{year}{2011}) \bibinfo{pages}{1--5}.
\bibitem[{Duan et~al.(2005)Duan, Prodan, and Fahringer}]{duan_dee:_2005}
\bibinfo{author}{R.~Duan}, \bibinfo{author}{R.~Prodan},
  \bibinfo{author}{T.~Fahringer},
\newblock \bibinfo{title}{{DEE:} a distributed fault tolerant workflow
  enactment engine for grid computing},
\newblock in: \bibinfo{booktitle}{First International Conference on High
  Performance Computing and Communications (HPCC 2005)}, volume
  \bibinfo{volume}{3726} of \textit{\bibinfo{series}{LNCS}},
  \bibinfo{publisher}{Springer}, \bibinfo{year}{2005}, pp.
  \bibinfo{pages}{704--716}.
\bibitem[{Liu et~al.(2010)Liu, Yuan, Zhang, Chen, and
  Yang}]{liu_swindew-c:_2010}
\bibinfo{author}{X.~Liu}, \bibinfo{author}{D.~Yuan},
  \bibinfo{author}{G.~Zhang}, \bibinfo{author}{J.~Chen},
  \bibinfo{author}{Y.~Yang},
\newblock \bibinfo{title}{{SwinDeW-C:} a peer-to-peer based cloud workflow
  system},
\newblock in: \bibinfo{editor}{B.~Furht}, \bibinfo{editor}{A.~Escalante}
  (Eds.), \bibinfo{booktitle}{Handbook of Cloud Computing},
  \bibinfo{publisher}{Springer}, \bibinfo{year}{2010}, pp.
  \bibinfo{pages}{309--332}.
\bibitem[{Stojni\'{c} and Schuldt(2013)}]{stojnic13}
\bibinfo{author}{N.~Stojni\'{c}}, \bibinfo{author}{H.~Schuldt},
\newblock \bibinfo{title}{{OSIRIS-SR: A Scalable Yet Reliable Distributed
  Workflow Execution Engine}},
\newblock in: \bibinfo{booktitle}{2nd ACM SIGMOD Workshop on Scalable Workflow
  Execution Engines and Technologies (SWEET '13)}, \bibinfo{publisher}{ACM},
  \bibinfo{year}{2013}, pp. \bibinfo{pages}{3:1--3:12}.
\bibitem[{Pantazoglou et~al.(2014)Pantazoglou, Pogkas, and
  Tsalgatidou}]{pantazoglou_decentralized_2013}
\bibinfo{author}{M.~Pantazoglou}, \bibinfo{author}{I.~Pogkas},
  \bibinfo{author}{A.~Tsalgatidou},
\newblock \bibinfo{title}{Decentralized enactment of {BPEL} processes},
\newblock \bibinfo{journal}{{IEEE} Transactions on Services Computing}
  \bibinfo{volume}{7} (\bibinfo{year}{2014}).
\bibitem[{Li et~al.(2010)Li, Muthusamy, and Jacobsen}]{li_distributed_2010}
\bibinfo{author}{G.~Li}, \bibinfo{author}{V.~Muthusamy}, \bibinfo{author}{H.-A.
  Jacobsen},
\newblock \bibinfo{title}{A distributed service-oriented architecture for
  business process execution},
\newblock \bibinfo{journal}{{ACM} Transactions on the Web} \bibinfo{volume}{4}
  (\bibinfo{year}{2010}) \bibinfo{pages}{2:1 -- 2:33}.
\bibitem[{Cheung and Jacobsen(2010)}]{cheung_load_2010}
\bibinfo{author}{A.~K.~Y. Cheung}, \bibinfo{author}{H.-A. Jacobsen},
\newblock \bibinfo{title}{Load balancing content-based {Publish/Subscribe}
  systems},
\newblock \bibinfo{journal}{{ACM} Transactions on Computer Systems}
  \bibinfo{volume}{28} (\bibinfo{year}{2010}) \bibinfo{pages}{9:1--9:55}.
\bibitem[{Benatallah et~al.(2005)Benatallah, Dumas, and
  Sheng}]{benatallah_facilitating_2005}
\bibinfo{author}{B.~Benatallah}, \bibinfo{author}{M.~Dumas},
  \bibinfo{author}{Q.~Z. Sheng},
\newblock \bibinfo{title}{Facilitating the rapid development and scalable
  orchestration of composite web services},
\newblock \bibinfo{journal}{Distributed and Parallel Databases}
  \bibinfo{volume}{17} (\bibinfo{year}{2005}) \bibinfo{pages}{5--37}.
\bibitem[{Islam et~al.(2012)Islam, Huang, Battisha, Chiang, Srinivasan, Peters,
  Neumann, and Abdelnur}]{islam_oozie:_2012}
\bibinfo{author}{M.~Islam}, \bibinfo{author}{A.~K. Huang},
  \bibinfo{author}{M.~Battisha}, \bibinfo{author}{M.~Chiang},
  \bibinfo{author}{S.~Srinivasan}, \bibinfo{author}{C.~Peters},
  \bibinfo{author}{A.~Neumann}, \bibinfo{author}{A.~Abdelnur},
\newblock \bibinfo{title}{Oozie: Towards a scalable workflow management system
  for hadoop},
\newblock in: \bibinfo{booktitle}{1st ACM SIGMOD Workshop on Scalable Workflow
  Execution Engines and Technologies (SWEET '12)}, \bibinfo{publisher}{ACM},
  \bibinfo{year}{2012}, pp. \bibinfo{pages}{4:1--4:10}.
\bibitem[{Warneke and Kao(2011)}]{warneke_exploiting_2011}
\bibinfo{author}{D.~Warneke}, \bibinfo{author}{O.~Kao},
\newblock \bibinfo{title}{Exploiting dynamic resource allocation for efficient
  parallel data processing in the cloud},
\newblock \bibinfo{journal}{{IEEE} Transactions on Parallel and Distributed
  Systems} \bibinfo{volume}{22} (\bibinfo{year}{2011})
  \bibinfo{pages}{985--997}.
\bibitem[{Muthusamy and Jacobsen(2010)}]{muthusamy_bpm_2010}
\bibinfo{author}{V.~Muthusamy}, \bibinfo{author}{H.-A. Jacobsen},
\newblock \bibinfo{title}{{BPM} in cloud architectures: Business process
  management with {SLAs} and events},
\newblock in: \bibinfo{booktitle}{8th International Conference on Business
  Process Management (BPM 2010)}, volume \bibinfo{volume}{6336} of
  \textit{\bibinfo{series}{LNCS}}, \bibinfo{publisher}{Springer},
  \bibinfo{year}{2010}, pp. \bibinfo{pages}{5--10}.
\bibitem[{Buchmann et~al.(2012)Buchmann, Appel, Freudenreich, Frischbier, and
  Guerrero}]{buchmann_calls_2012}
\bibinfo{author}{A.~Buchmann}, \bibinfo{author}{S.~Appel},
  \bibinfo{author}{T.~Freudenreich}, \bibinfo{author}{S.~Frischbier},
  \bibinfo{author}{P.~E. Guerrero},
\newblock \bibinfo{title}{From calls to events: Architecting future {BPM}
  systems},
\newblock in: \bibinfo{booktitle}{10th International Conference on Business
  Process Management (BPM 2012)}, volume \bibinfo{volume}{7481} of
  \textit{\bibinfo{series}{LNCS}}, \bibinfo{publisher}{Springer},
  \bibinfo{year}{2012}, pp. \bibinfo{pages}{17--32}.
\bibitem[{Schuster et~al.(1999)Schuster, Neeb, and
  Schamburger}]{schuster_configuration_1999}
\bibinfo{author}{H.~Schuster}, \bibinfo{author}{J.~Neeb},
  \bibinfo{author}{R.~Schamburger},
\newblock \bibinfo{title}{A configuration management approach for large
  workflow management systems},
\newblock in: \bibinfo{booktitle}{International Joint Conference on Work
  Activities Coordination and Collaboration ({WACC} '99)},
  \bibinfo{publisher}{{ACM}}, \bibinfo{year}{1999}, pp.
  \bibinfo{pages}{177--186}.
\bibitem[{Rinderle et~al.(2004)Rinderle, Reichert, and
  Dadam}]{rinderle_correctness_2004}
\bibinfo{author}{S.~Rinderle}, \bibinfo{author}{M.~Reichert},
  \bibinfo{author}{P.~Dadam},
\newblock \bibinfo{title}{Correctness criteria for dynamic changes in workflow
  systems - a survey},
\newblock \bibinfo{journal}{Data \& Knowledge Engineering} \bibinfo{volume}{50}
  (\bibinfo{year}{2004}) \bibinfo{pages}{9--34}.
\bibitem[{Han et~al.(1998)Han, Sheth, and Bussler}]{han_taxonomy_1998}
\bibinfo{author}{Y.~Han}, \bibinfo{author}{A.~Sheth},
  \bibinfo{author}{C.~Bussler},
\newblock \bibinfo{title}{A taxonomy of adaptive workflow management},
\newblock in: \bibinfo{booktitle}{Workshop of the 1998 {ACM} Conference on
  Computer-Supported Cooperative Work (CSCW-98)}, \bibinfo{publisher}{ACM},
  \bibinfo{year}{1998}.
\bibitem[{Reichert and Dadam(1998)}]{reichert_adeptflexsupporting_1998}
\bibinfo{author}{M.~Reichert}, \bibinfo{author}{P.~Dadam},
\newblock \bibinfo{title}{Adeptflex - supporting dynamic changes of workflows
  without losing control},
\newblock \bibinfo{journal}{Journal of Intelligent Information Systems}
  \bibinfo{volume}{10} (\bibinfo{year}{1998}) \bibinfo{pages}{93--129}.
\bibitem[{Burgess and Fagernes(2006)}]{burgess_promise_2006}
\bibinfo{author}{M.~Burgess}, \bibinfo{author}{S.~Fagernes},
\newblock \bibinfo{title}{Promise theory -- a model of autonomous objects for
  pervasive computing and swarms},
\newblock in: \bibinfo{booktitle}{International Conference on Networking and
  Services ({ICNS}'06)}, \bibinfo{publisher}{IEEE}, \bibinfo{year}{2006}, p.
  \bibinfo{pages}{118}.
\bibitem[{Xu et~al.(2012)Xu, Rao, and Bu}]{xu_url:_2012}
\bibinfo{author}{C.-Z. Xu}, \bibinfo{author}{J.~Rao}, \bibinfo{author}{X.~Bu},
\newblock \bibinfo{title}{{URL:} a unified reinforcement learning approach for
  autonomic cloud management},
\newblock \bibinfo{journal}{Journal of Parallel and Distributed Computing}
  \bibinfo{volume}{72} (\bibinfo{year}{2012}) \bibinfo{pages}{95--105}.
\bibitem[{Wei and Blake(2013)}]{wei13a}
\bibinfo{author}{Y.~Wei}, \bibinfo{author}{M.~B. Blake},
\newblock \bibinfo{title}{{Adaptive Service Workflow Configuration and
  Agent-Based Virtual Resource Management in the Cloud}},
\newblock in: \bibinfo{booktitle}{2013 IEEE International Conference on Cloud
  Engineering (IC2E 2013)}, \bibinfo{publisher}{IEEE}, \bibinfo{year}{2013},
  pp. \bibinfo{pages}{279--284}.
\bibitem[{Sakr(2014)}]{sakr_cloud-hosted_2014}
\bibinfo{author}{S.~Sakr},
\newblock \bibinfo{title}{Cloud-hosted databases: technologies, challenges and
  opportunities},
\newblock \bibinfo{journal}{Cluster Computing} \bibinfo{volume}{17}
  (\bibinfo{year}{2014}) \bibinfo{pages}{487--502}.
\bibitem[{Li and Venugopal(2013)}]{li_efficient_2013}
\bibinfo{author}{H.~Li}, \bibinfo{author}{S.~Venugopal},
\newblock \bibinfo{title}{Efficient node bootstrapping for decentralised
  shared-nothing key-value stores},
\newblock in: \bibinfo{booktitle}{ACM/IFIP/USENIX 14th International Middleware
  Conference (Middleware 2013)}, volume \bibinfo{volume}{8275} of
  \textit{\bibinfo{series}{LNCS}}, \bibinfo{publisher}{Springer},
  \bibinfo{year}{2013}, pp. \bibinfo{pages}{348--367}.
\bibitem[{Mangler et~al.(2010)Mangler, St\"{u}rmer, and Schikuta}]{mangler10}
\bibinfo{author}{J.~Mangler}, \bibinfo{author}{G.~St\"{u}rmer},
  \bibinfo{author}{E.~Schikuta},
\newblock \bibinfo{title}{{Cloud Process Execution Engine -- Evaluation of the
  Core Concepts}},
\newblock \bibinfo{journal}{CoRR} \bibinfo{volume}{abs/1003.3330}
  (\bibinfo{year}{2010}).
\bibitem[{Mamdani(1974)}]{Mamdani1974}
\bibinfo{author}{E.~H. Mamdani},
\newblock \bibinfo{title}{{Application of Fuzzy Algorithm for Control of Simple
  Dynamic Plant}},
\newblock \bibinfo{journal}{Proceedings of the IEEE} \bibinfo{volume}{121}
  (\bibinfo{year}{1974}) \bibinfo{pages}{1585--1588}.
\bibitem[{Schulte et~al.(2013)Schulte, Hoenisch, Venugopal, and
  Dustdar}]{schulte13}
\bibinfo{author}{S.~Schulte}, \bibinfo{author}{P.~Hoenisch},
  \bibinfo{author}{S.~Venugopal}, \bibinfo{author}{S.~Dustdar},
\newblock \bibinfo{title}{{Introducing the Vienna Platform for Elastic
  Processes}},
\newblock in: \bibinfo{booktitle}{Performance Assessment and Auditing in
  Service Computing Workshop (PAASC 2012) at 10th International Conference on
  Service Oriented Computing (ICSOC 2012)}, volume \bibinfo{volume}{7759} of
  \textit{\bibinfo{series}{LNCS}}, \bibinfo{publisher}{Springer},
  \bibinfo{year}{2013}, pp. \bibinfo{pages}{179--190}.
\bibitem[{{Workflow Management Coalition}(2008)}]{wfmc08}
\bibinfo{author}{{Workflow Management Coalition}}, \bibinfo{title}{Business
  process analytics format, draft version 2.0}, \bibinfo{year}{2008}.
\bibitem[{{Workflow Management Coalition}(2014)}]{guentherverbeek14}
\bibinfo{author}{{Workflow Management Coalition}}, \bibinfo{title}{{XES}
  standard definition. version 2.0}, \bibinfo{year}{2014}.
\bibitem[{{W. M. P. van der Aalst et al.}(2011)}]{aalst11}
\bibinfo{author}{{W. M. P. van der Aalst et al.}},
\newblock \bibinfo{title}{{Process Mining Manifesto}},
\newblock in: \bibinfo{booktitle}{Business Process Management Workshops - BPM
  2011 International Workshops, Revised Selected Papers, Part I},
  volume~\bibinfo{volume}{99} of \textit{\bibinfo{series}{LNBIP}},
  \bibinfo{publisher}{Springer}, \bibinfo{year}{2011}, pp.
  \bibinfo{pages}{169--194}.
\bibitem[{{Workflow Management Coalition}(1998)}]{wfmc98}
\bibinfo{author}{{Workflow Management Coalition}}, \bibinfo{title}{{Workflow
  Management Application Programming Interface (Inteface 2\&3) Specification,
  Version 2.0}}, \bibinfo{year}{1998}.

\end{thebibliography}
\end{document}